\documentclass[12pt,draftcls,onecolumn]{IEEEtran}

\ifCLASSINFOpdf
\else
\fi

\ifCLASSOPTIONcompsoc
\usepackage[nocompress]{cite}
\else
\usepackage{cite}
\fi

\usepackage{amsmath}
\usepackage{amssymb}
\usepackage{graphicx}
\usepackage{epstopdf}
\usepackage{mathtools}
\usepackage{xcolor}

\begin{document}

\title{Omnidirectional Space-Time Block Coding for Common Information Broadcasting in \\ Massive MIMO Systems}

\author{Xin Meng, Xiang-Gen Xia, and Xiqi Gao
\thanks{X. Meng and X. Q. Gao are with the National Mobile Communications Research Laboratory, Southeast University, Nanjing, 210096, China (e-mail: \{xmeng, xqgao\}@seu.edu.cn).}
\thanks{X.-G. Xia is with the Department of Electrical and Computer Engineering, University of Delaware, Newark, DE 19716, USA (e-mail: xxia@ee.udel.edu).}
}



\maketitle

\vspace*{-20pt}

\begin{abstract}
In this paper, we design space-time block codes (STBCs) to broadcast the common information omnidirectionally in a massive MIMO downlink. To reduce the burden of the downlink channel estimation and achieve partial spatial diversity from base station (BS) transmit antennas, we propose channel-independently precoded low-dimensional STBC. The precoding matrix and the signal constellation in the low-dimensional STBC are jointly designed to guarantee omnidirectional coverage at any instant time and sufficiently utilize the power amplifier capacities of BS  transmit antennas, and at the same time, achieve the full diversity of the low-dimensional STBC. Under this framework, several designs are presented. To provide transmit diversity order of two, a precoded Alamouti code is proposed, which has a fast symbol-wise maximum-likelihood (ML) decoding. To provide transmit diversity order of four,  three types of STBCs are proposed, being referred to as precoded orthogonal STBC (OSTBC), precoded quasi-orthogonal STBC (QOSTBC), and precoded coordinate interleaved orthogonal design (CIOD), respectively. The last two codes have the same complexity for pair-wise ML decoding, while precoded QOSTBC has a higher coding gain when the bit rate is lower than or equal to 4 bps/Hz, and precoded CIOD has a higher coding gain when the bit rate is higher than 4 bps/Hz. Precoded OSTBC has a higher decoding complexity and a lower coding gain than the other two codes, since in the precoded OSTBC the information symbols need to be jointly designed and decoded. Moreover, a precoded no-zero-entry Toeplitz code and a precoded no-zero-entry overlapped Alamouti code are also proposed. These two codes can achieve a higher diversity order with linear receivers.
\end{abstract}



\section{Introduction}

Massive multiple-input multiple-output (MIMO) has received considerable interest from both academia and industry in recent years, which is regarded as a key technique in the fifth generation (5G) of cellular wireless communication systems \cite{T.L.Marzetta2010,L.Lu2014,E.G.Larsson2014,C.-X.Wang2014,J.G.Andrews2014}. Owing to the deployment of a large number of antennas at the base station (BS) side, massive MIMO systems are expected to increase the energy and spectral efficiency significantly even with simple linear signal processing \cite{H.Q.Ngo2013,F.Rusek2013}. These advantages are mainly harvested by serving tens of active user terminals (UTs) on the same time-frequency resource simultaneously with spatially directional narrow beams formed by the BS's massive antenna array, which leads to a high power gain for each UT and a high multiplexing gain for the entire system. Moreover, the energy-focusing effect provided by narrow beams can also improve the physical layer security \cite{D.Kapetanovic2015,Y.Wu2016}.

Most of the previous researches on the massive MIMO downlink have been focused on the case where the BS transmits individual information to different UTs, which is also known as broadcast channel or multi-user downlink transmission. Besides, it is also of great interest and plays an important role in cellular systems, where the BS broadcasts the same common information to multiple UTs simultaneously. This is also known as multicasting in some literature. A typical scenario is that the BS broadcasts the common control signaling to activate a ``sleeping'' UT or delivers popular audio/video data to a group of subscription UTs. Wireless common information broadcasting has also been included in the third generation partnership project (3GPP) long-term evolution (LTE) standards known as evolved multimedia broadcast multicast service (eMBMS) \cite{D.Lecompte2012}.

Note that common information broadcasting has been investigated in many previous studies. According to whether the BS utilizes the channel state information (CSI) or not, the approaches for common information broadcasting can be divided into two categories: closed-loop and open-loop. For closed-loop approaches, it is typically assumed that the instantaneous or statistic CSI between the BS and the UTs is known at the BS side. Then by exploiting the CSI, the BS determines the corresponding transmission strategy, e.g., choosing the optimal precoding matrix to maximize the worst-case receiving signal-to-noise ratio (SNR) \cite{Y.Sun2004,J.Wang2006,N.D.Sidiropoulos2006,A.Lozano2007,Z.Xiang2014}. For open-loop approaches, the BS does not utilize any CSI and needs to broadcast the common information blindly regardless of the UTs. A typical open-loop approach is to use space-time block codes (STBCs), e.g., the Alamouti code (AC) which is suitable for two transmit antennas \cite{S.M.Alamouti1998}. In this paper, we focus on the open-loop approach. We restrict our concern to STBC transmissions in massive MIMO systems for common information broadcasting.

One of the major concerns about STBC transmissions is instantaneous CSI acquisition at the receiver side, with which the transmitted codeword can be decoded coherently. A typical method is to send pilots at the transmitter side for channel estimation at the receiver side. To obtain a meaningful estimation value, the length of the pilots should not be less than the number of transmit antennas. In a massive MIMO downlink where the BS (transmitter) has a large number of antennas, too many time-frequency resources would have to be spent on the pilots, hence lowering the net spectral efficiency to a great extent. To address this problem, the idea of confining the transmitted signal to lie in a low-dimensional subspace to reduce the pilot overhead was independently proposed in \cite{X.Meng2016,X.Meng2014,X.Meng2015,M.Karlsson2014}. The authors in \cite{M.Karlsson2014} considered the independent and identically distributed (i.i.d.) channel and proposed to repeat a low-dimensional signal across BS antennas to reduce the pilot overhead. The transmitted signal in this scheme will be spatially selective. For practical spatially correlated channels, the transmitted signal should be spatially omnidirectional, i.e., having equal radiation power in each spatial direction. Such that the UT in any spatial direction can obtain a fair receiving SNR. Besides omnidirectional transmission, it is also important for the transmitted signal to have equal power on each antenna to sufficiently utilize all the power amplifier (PA) capacities of the BS. In \cite{X.Meng2016}, we proposed to map a low-dimensional signal to the high-dimensional antenna array of the BS through a channel-independent precoding matrix to reduce the pilot overhead. The precoding matrix is specially designed to satisfy the above two power constraints statistically in a long time period. In this paper, the same with \cite{X.Meng2014,X.Meng2015}, we consider more strict power constraints where the transmission power is constant across spatial directions and transmit antennas at {\it any instant time} (not just in the statistics sense), and then design the precoding matrix and the signal constellation in the low-dimensional STBC jointly to satisfy the constraints.

The remainder of this paper is organized as follows. The system model is presented in Section II, including the channel model and the framework of precoding based STBC transmission. In Section III, two basic requirements that the STBC should satisfy are demonstrated. In Section IV, a systematic approach to design such an STBC and the corresponding diversity performance analysis are presented. In Section V, several detailed examples of the STBC are designed. Numerical results are presented in Section VI. Finally, conclusions are drawn in Section VII.

\emph{Notations:} We use upper-case and lower-case boldfaces to denote matrices and column vectors. Specifically, ${\bf{I}}_M$, ${\bf{F}}_M$, ${\bf{1}}_{M}$, and $\bf{0}$ denote the $M \times M$ identity matrix, the unitary $M$-point discrete Fourier transform (DFT) matrix, the $M \times 1$ column vector of all ones, and the zero matrix with proper dimensions, respectively. Let ${(\cdot)}^*$, ${(\cdot)}^T$, and ${(\cdot)}^H$ denote the conjugate, the transpose, and the conjugate transpose, respectively. ${{[ {\bf{A}} ]}_{m,n}}$ denotes the $( {m,n} )$th element of matrix $\bf{A}$ and ${[{\bf a}]}_m$ denotes the $m$th element of vector $\bf a$. ${\mathbb{E}}(\cdot)$ refers to the expectation and ${\mathbb{P}}(\cdot)$ represents the probability. The Kronecker product of two matrices $\bf{A}$ and $\bf{B}$ is denoted by ${\bf{A}} \otimes {\bf{B}}$. ${\rm{diag}}( {\bf{A}} )$ and ${\rm{diag}}( {\bf{a}} )$ denote the column vector constituted by the main diagonal of $\bf{A}$ and the diagonal matrix with $\bf{a}$ on the main diagonal, respectively. ${{((a))}_b}$ denotes $a$ modulo $b$.

\section{System Model}

\subsection{Channel Model}

Consider a single cell, where the BS is equipped with a uniform linear array (ULA) of $M$ antennas and serves $K$ UTs each with a single antenna. Assume Rayleigh flat-fading channel. Let ${\bf{h}}_k \in \mathbb{C}^{M \times 1}$ denote the channel vector between the BS and the $k$th UT, we have
\begin{align}
{\bf{h}}_k \sim \mathcal{CN}( {{\bf{0}},{\bf{R}}_k} ) . \label{Channel Model 0}
\end{align}
For the one-ring scattering model under a far-field assumption, the channel covariance matrix ${\bf{R}}_k$ is generated by \cite{A.Adhikary2013,Y.S.Cho2010}
\begin{align}\label{Channel Model 1}
{\bf{R}}_k = \int_{ - \pi /{2}}^{\pi /{2}} {{\bf{v}}( \theta  ) {({{\bf{v}}}( \theta  ))}^H p_k( \theta  )\mathrm{d}\theta }
\end{align}
where ${\bf{v}}( \theta  ) = {[1,{e^{-j2\pi d \sin \theta /\lambda }}, \ldots ,{e^{-j2\pi ( {M - 1} ) d \sin \theta /\lambda }}]}^T$ represents the steering vector of the ULA, $d$ represents the antenna space, $\lambda$ represents the carrier wavelength, and ${p_k( \theta  )}$ represents the power azimuth spectrum (PAS), which may follow different distributions, e.g., truncated Gaussian distribution or truncated Laplacian distribution, depending on the characteristics of the terrain \cite{Y.S.Cho2010}.

\emph{Lemma 1 \cite{A.Adhikary2013,S.Noh2014,U.Grenander1958,A.A.Lu2016,L.You2015,L.You2016}:} When the number of BS antennas $M$ is sufficiently large, the channel covariance matrix ${\bf R}_k$ defined in (\ref{Channel Model 1}) asymptotically has the eigenvalue decomposition
\begin{align}
{\bf{R}}_k \xrightarrow{M \to \infty } {\bf{F}}_M^H{\bf{\Lambda }}_k{{\bf{F}}_M} \nonumber
\end{align}
where ${\bf F}_M$ is the unitary $M$-point DFT matrix and ${\bf \Lambda}_k$ is a diagonal matrix with non-negative diagonal elements.

With Lemma 1, for a sufficiently large number of BS antennas in the massive MIMO regime, the channel covariance matrix ${\bf R}_k$ in (\ref{Channel Model 1}) can hence be well approximated by the asymptotic result, i.e., ${\bf{R}}_k \approx {\bf{F}}_M^H{\bf{\Lambda }}_k{{\bf{F}}_M}$. This approximation has been shown to be accurate enough with a practical value of the number of antennas which usually ranges from $64$ to $512$ \cite{C.K.Wen2015,H.Yin2013}. Therefore we assume
\begin{align}\label{Channel Model 2}
{\bf{h}}_k \sim \mathcal{CN}( {{\bf{0}},{\bf{F}}_M^H{\bf{\Lambda }}_k{{\bf{F}}_M}} )
\end{align}
as the basic channel model to simplify our analysis and designs. In simulations, we will still use the non-asymptotic model (\ref{Channel Model 1}) to generate the channel covariance matrix to evaluate the performance for our designs. Moreover, without loss of generality, these $K$ UTs are assumed to experience the same large-scale fading, i.e., $\mathrm{tr}( {{{\bf{R }}_k}} ) = \mathrm{tr}( {{{\bf{\Lambda }}_k}} ) = M$ for $k = 1,2,\ldots,K$.

\subsection{Precoding Based STBC Transmission}

Consider STBC transmission for common information broadcasting. Assume that the common information, which can be regarded as a group of binary bits, is mapped to an STBC matrix ${\bf{S}} \in \mathbb{C}^{M \times T}$. This codeword matrix is then transmitted from the $M$ antennas of the BS within $T$ time slots. The received signal at the $k$th UT follows
\begin{align}\label{System Model 1}
[{y_{k,1}},{y_{k,2}}, \ldots ,{y_{k,T}}] = {\bf{h}}_k^H{\bf{S}} + [{z_{k,1}},{z_{k,2}}, \ldots ,{z_{k,T}}]
\end{align}
where the channel ${{\bf{h}}}_k$ is assumed to keep constant within these $T$ time slots, and $z_{k,t} \sim \mathcal{CN}( {0,\sigma ^2_{\mathrm{n}}} )$ for $t = 1,2,\ldots,T$ denotes the additive white Gaussian noise (AWGN).

With signal model (\ref{System Model 1}), to coherently decode the transmitted codeword $\bf S$, the instantaneous CSI ${\bf{h}}_k$ with dimension $M$ must be known at the UT side. When utilizing training based downlink channel estimation, the length of the downlink pilots should not be less than $M$. In a massive MIMO system where the number of BS antennas $M$ is large, many time-frequency resources would have to be spent on the pilots. This will lower the net spectral efficiency to a large extent.

In order to reduce the pilot overhead, we propose that the high-dimensional STBC is composed by a precoding matrix and a low-dimensional STBC. Correspondingly, we can write (\ref{System Model 1}) as
\begin{align}\label{System Model 2}
{[y_{k,1} , y_{k,2},\ldots , y_{k,T} ]} = {{\bf{h}}_k^H}{\bf{W}}{{\bf{X}}} + {[z_{k,1} , z_{k,2},\ldots , z_{k,T} ]}
\end{align}
where ${\bf{W}} \in \mathbb{C}^{M \times N}$ is a tall precoding matrix since $N$ is selected to be less than $M$, and ${{\bf{X}}} \in \mathbb{C}^{N \times T}$ is a low-dimensional STBC. With (\ref{System Model 2}), the UT does not need to estimate the actual channel ${\bf{h}}_k$, but only requires to estimate the effective channel ${{\bf{W}}^H}{\bf{h}}_k$, and then decodes the codeword $\bf X$. Therefore, the length of the downlink pilots can be reduced to $N$, which is the dimension of ${{\bf{W}}^H}{\bf{h}}_k$.  As long as $N$ is selected to be sufficiently small, the pilot overhead can also be reduced to small enough.

Note that the main purpose of using the precoding matrix in (\ref{System Model 2}) is different from those in \cite{Y.Sun2004,J.Wang2006,N.D.Sidiropoulos2006,A.Lozano2007,Z.Xiang2014}, where the precoding matrices depend on either the instantaneous CSI or the statistical CSI, and are used to improve the performance with spatially directional signaling. However, the precoding matrix here is channel-independent and mainly used to ease channel estimation. Signal model (\ref{System Model 2}) is the basic framework in this paper. In what follows, we will mainly discuss how to design the precoding matrix $\bf W$ and the low-dimensional STBC $\bf X$, for common information broadcasting, and in the meantime, satisfying the requirements on power efficiency and diversity performance. Moreover, we assume that $\mathbb{E}({\bf X}{\bf X}^H) = T \cdot {\bf I}_N$ and $\mathrm{tr} ( {\bf W}{\bf W}^H ) = 1$ to normalize the total average transmission power at the BS side. Before proceeding, we will discuss the basic requirements that the STBC should satisfy in the next section.

\section{Basic Requirements of the STBC}

\subsection{Omnidirectional Transmission}

In signal model (\ref{System Model 2}), letting ${\bf x}_t$ represent the $t$th column of $\bf X$, we can denote ${\bf W}{\bf{x}}_t $ as the transmitted signal vector from the $M$ antennas of the BS at time slot $t$. Then, ${{\bf{h}}_k^H}{\bf{W}}{{\bf{x}}_t}$ is the received signal at the $k$th UT without AWGN. Hence, the power of the received signal can be expressed as ${| {{{\bf{h}}_k^H}{\bf W}{{\bf{x}}_t}} |}^2$. For common information broadcasting, it is expected that the BS can provide equal receiving power for all the UTs, hence all the UTs can have fair quality-of-services (QoS). In mathematical expressions, it is expected that ${| {{{\bf{h}}_k^H}{\bf W}{{\bf{x}}_t}} |}^2$ is constant regardless of ${\bf h}_k \neq {\bf 0}$. Unfortunately, this will be impossible by only designing ${\bf W}$ and ${{\bf{x}}_t}$ when ${\bf h}_k$ is unknown at the BS side. One reasonable method is to average over the fast fading of ${\bf h}_k$. With (\ref{Channel Model 2}), the received power after this averaging can hence be expressed as
\begin{align}\label{Average Receiving Power}
P({\bf \Lambda}_k) = \mathbb{E}\big\{ {{{| {{{\bf{h}}_k^H}{\bf W}{{\bf{x}}_t}} |}^2}} \big\} = {\bf{x}}_t^H{{\bf{W}}^H}{\bf{F}}_M^H{\bf{\Lambda }}_k{{\bf{F}}_M}{\bf{W}}{{\bf{x}}_t}
\end{align}
where ${{\bf{W}}^H}{\bf{F}}_M^H{\bf{\Lambda }}_k{{\bf{F}}_M}{\bf{W}}$ is the covariance matrix of the effective channel ${\bf W}^H {\bf h}_k$. However, without knowing ${\bf \Lambda}_k$ in (\ref{Average Receiving Power}), it is still impossible to let $P({\bf \Lambda}_k)$ be constant regardless of ${\bf \Lambda}_k \neq {\bf 0}$ by only designing ${\bf W}$ and ${{\bf{x}}_t}$. Therefore, we restrict our concern to the case that all the $K$ UTs have the same large-scale fading with respect to the BS, i.e., letting the diagonal matrix ${\bf{\Lambda }}_k \in \mathcal{A}$ for $k=1,2,\ldots,K$ where $\mathcal{A} = \{ {{\bf{\Lambda }}|{\rm{tr}}( {\bf{\Lambda }} ) = M} \}$. Then we have the following lemma.

\emph{Lemma 2:} For any diagonal matrix ${\bf{\Lambda }} \in \mathcal{A}$, the average receiving power $P({\bf \Lambda})$ is constant if and only if all the $M$ elements in ${\bf{F}}_M{\bf W}{{\bf{x}}_t}$ have the same amplitude.
\begin{IEEEproof}
For notational simplicity, let ${\bf a} = {\bf{F}}_M{\bf W}{{\bf{x}}_t} $, where the $m$th element of $\bf a$ is denoted by $a_m$ for $m=1,2,\ldots,M$, and let $\lambda_m$ be the $m$th element of the main diagonal of ${\bf \Lambda}$. When all the $M$ elements in $\bf a$ have the same amplitude, i.e., $|{a_1}| = |{a_2}| =  \cdots  = |{a_M}| = a$, we have
\begin{align}\nonumber
{{\bf{a}}^H}{\bf{\Lambda a}} = \sum\limits_{m = 1}^M {{|{a_m}|}^2{\lambda _m}}  = a^2 \cdot {\rm{tr}}({\bf{\Lambda }}) = M a^2
\end{align}
for any diagonal matrix ${\bf{\Lambda }} \in \mathcal{A}$.  This verifies the sufficiency. Then we prove the necessity. If all the $M$ elements in $\bf a$ do not have the same amplitude, at least two of them do not have the same amplitude. Without loss of generality, we can assume that $|a_1| \neq |a_2|$. For both of the following two diagonal matrices ${{\bf{\Lambda }}_1} = {\rm{diag}}\{ M,0,0, \ldots ,0\} $ and ${{\bf{\Lambda }}_2} = {\rm{diag}}\{ 0,M,0, \ldots ,0\} $ belonging to $\mathcal{A}$, it can be shown that ${{\bf{a}}^H}{{\bf{\Lambda }}_1}{\bf{a}} = M{|{a_1}|}^2  $ and ${{\bf{a}}^H}{{\bf{\Lambda }}_2}{\bf{a}} = M{|{a_2}|}^2  $. Hence ${{\bf{a}}^H}{{\bf{\Lambda }}_1}{\bf{a}} \neq {{\bf{a}}^H}{{\bf{\Lambda }}_2}{\bf{a}}$ since $|a_1| \neq |a_2|$. This verifies the necessity.
\end{IEEEproof}

The $M$ elements of ${\bf{F}}_M{\bf W}{{\bf{x}}_t}$ can be seen as the transmitted signals in $M$ discrete spatial directions, respectively. Correspondingly, the squared absolute values of these $M$ elements represent the transmission power in $M$ discrete spatial directions, respectively.  Hence, all the $M$ elements in ${\bf{F}}_M{\bf W}{{\bf{x}}_t}$ having the same squared absolute value, i.e., the same amplitude, means that the transmitted signal has equal power in all discrete spatial directions, i.e., radiating omnidirectionally. Then we have the following requirement.

\emph{Requirement 1:} To have omnidirectional transmission, all the $M$ elements in ${\bf{F}}_M{\bf W}{{\bf{x}}_t}$ should have the same amplitude, where ${\bf W}{{\bf{x}}_t}$ is the transmitted signal vector from the $M$ transmit antennas at time slot $t$.

\subsection{Equal Power on Each Antenna}

Besides omnidirectional transmission, the PA utilization efficiency at transmit antennas also needs to be considered. In practice, each transmit antenna has its own PA in its analog front-end, and is limited individually by the linearity of the PA. Therefore, the transmission power on each antenna cannot exceed a maximum value for that antenna. We assume that the maximum power value on each antenna is equal to each other. This is reasonable when all the antennas employ the same PA hardware. Let $s_{m,t} = {[{\bf W}{\bf x}_t]}_m$ denote the transmitted signal on the $m$th antenna at time slot $t$. Per-antenna power constraint means that ${|s_{m,t}|}^2 \leq P $ for $m=1,2,\ldots,M$ and any $t$, where $P$ denotes the maximum allowable power on each antenna. At the same time, to utilize the maximum PA capacity of BS antennas, all these $M$ antennas need to transmit with the maximum power $P$, i.e., ${|s_{1,t}|}^2 = {|s_{2,t}|}^2 = \cdots = {|s_{M,t}|}^2 = P$. Furthermore, if ${|s_{1,t}|}^2 = {|s_{2,t}|}^2 = \cdots = {|s_{M,t}|}^2 = c \neq P$, we can always multiply each ${s}_{m,t}$ with a constant scalar $\sqrt{P/c}$ without causing any distortion to the signal. Then all the $M$ antennas have transmission power $c\cdot P/c = P$. Therefore, we have the following requirement.

\emph{Requirement 2:} To have equal power on each transmit antenna to sufficiently utilize all the PA capacities of BS antennas, all the $M$ elements in ${\bf W}{{\bf{x}}_t}$ should have the same amplitude.

Note that in this paper, equal instantaneous power across spatial directions and transmit antennas is considered, i.e., at each time slot $t$, the instantaneous transmission power keeps constant for all the spatial directions and all the transmit antennas. In our previous paper \cite{X.Meng2016}, equal average power across spatial directions and transmit antennas is considered, i.e., in a long time period, the average transmission power keeps constant for all the spatial directions and all the transmit antennas, where the average is taken over the information symbols in the low-dimensional STBC $\bf X$ in (\ref{System Model 2}). The average power just depends on the precoding matrix $\bf{W}$ since the information symbols have been averaged out, while the instantaneous power depends on both the precoding matrix $\bf W$ and the information symbols in the low-dimensional STBC $\bf X$. This implies that a much more strict power constraint is considered in this paper. Therefore, the signal constellation in the low-dimensional STBC needs to be jointly designed with the precoding matrix to satisfy the above two requirements, while any existing  low-dimensional STBC can be directly employed in \cite{X.Meng2016}.

\section{Main Framework for STBC Designs}

In this section, we propose a systematic approach to design STBCs satisfying Requirements 1--2 in Section III, and then analyze the diversity performance for the STBCs designed with the proposed approach.

\subsection{The Approach to Design STBCs}

First, some useful mathematical results are presented to help the STBC design. We refer to a sequence of length $M$ as a constant-amplitude zero auto-correlation (CAZAC) sequence if all the $M$ elements of this sequence have the same amplitude, and at the same time, all the $M$ elements of the $M$-point DFT of this sequence have the same amplitude. Zadoff-Chu (ZC) sequences are well-known CAZAC sequences. A ZC sequence ${\bf c} = \frac{1}{\sqrt{M}}\cdot{[c_0,c_1,\ldots,c_{M-1}]}^T $ of length $M$  is defined as\footnote{Throughout this paper, we use an $M\times 1$ column vector to denote a sequence of length $M$.} \cite{D.Chu1972}
\begin{align}
{c}_m = \left\{ {\begin{matrix*}[l]
{{e^{j\pi \gamma {m^2}/M} },}&{M \text{ is even}}\\
{{e^{j\pi \gamma m( {m + 1} )/M}},}&{M \text{ is odd}}
\end{matrix*}} \right. \label{ZC Sequence}
\end{align}
where the parameter $\gamma$, known as the root of the ZC sequence, is an integer less than and relatively prime to $M$. With (\ref{ZC Sequence}), one can see that all the $M$ elements in $\bf c$ have the same amplitude. Furthermore, it is shown in \cite{D.Chu1972} that the ZC sequence has perfect periodic auto-correlation whether $M$ is even or odd, i.e.,
\begin{align}
{{\bf{c }}^H}{{\bf{\Pi }}_n}{\bf{c }} = \delta _n \nonumber
\end{align}
where $\delta _n$ denotes the Kronecker delta function and
\begin{align}
{{\bf{\Pi }}_n} = {\begin{bmatrix}
{{{\bf{0}}}}&{{{\bf{I}}_n}}\\
{{{\bf{I}}_{M - n}}}&{{{\bf{0}}}}
\end{bmatrix}} \nonumber
\end{align}
for $n = 0,1,\ldots,M-1$ denotes the cyclic shifting matrix. Since the periodic auto-correlation function of $\bf{c}$ is a Kronecker delta function, with the fact that periodic auto-correlation function and power spectrum function are DFT pairs, the power spectrum function of $\bf c$, denoted by an $M \times 1$ column vector $\mathrm{diag} ( {\bf{F}}_M {\bf{c}} {\bf{c}}^H {\bf{F}}_M^H ) $, will have $M$ identical elements due to the DFT of a Kronecker delta function. Therefore, all the $M$ elements of ${\bf{F}}_M {\bf{c}}$ have the same amplitude whether $M$ is even or odd.

Based on the ZC sequence, we introduce the following lemma which will be used latter.

\emph{Lemma 3:} Consider that $N$ is an integer and $M$ is an integer multiple of $N^2$. Let $\bf x$ be an $N \times 1$ vector and ${\rm{diag}}( {\bf{c}} )( {{{\bf{1}}_{M/N}} \otimes {\bf{x}}} )$ be an $M \times 1$ vector constructed from $\bf x$, where $\bf{c}$ is a ZC sequence of length $M$ defined in (\ref{ZC Sequence}), ${\bf{1}}_{M/N}$ is a column vector with $M/N$ ones, and $\otimes$ denotes the Kronecker product. If and only if all the $N$ elements in ${\bf{x}} $ have the same amplitude, ${\rm{diag}}( {\bf{c}} )( {{{\bf{1}}_{M/N}} \otimes {\bf{x}}} )$ is a CAZAC sequence.
\begin{IEEEproof}
See Appendix A.
\end{IEEEproof}

With the above results, we propose a systematic approach to design such an STBC satisfying Requirements 1--2 in Section III. Consider a low-dimensional STBC ${\bf X} $ of size $N \times T$ and the following precoding matrix
\begin{align}
{\bf{W}} = {\rm{diag}}( {\bf{c}} )( {{{\bf{1}}_{M/N}} \otimes {{\bf{V}}}} ) \label{General Precoding Matrix}
\end{align}
of size $M \times N$, where $\bf{c}$ is a ZC sequence of length $M$ defined in (\ref{ZC Sequence}), ${\bf{1}}_{M/N}$ is a column vector with $\frac{M}{N}$ ones, $\otimes$ denotes the Kronecker product, and $\bf V$ is an $N \times N$ unitary matrix. Then the transmitted signals on the $M$ transmit antennas over $T$ time slots are given by
\begin{align}
{\bf{WX}} &= {\rm{diag}}( {\bf{c}} )( {{{\bf{1}}_{M/N}} \otimes {{\bf{V}}}} ) {\bf X} \nonumber \\
&= {\rm{diag}}( {\bf{c}} )( {{{\bf{1}}_{M/N}} \otimes {{\bf{V}}}} ) (1 \otimes {\bf X}) \nonumber \\
&= {\rm{diag}}( {\bf{c}} )( {{{\bf{1}}_{M/N}} \otimes ({{\bf{VX}}})} ) \nonumber
\end{align}
where the last equation is with $({\bf A}\otimes{\bf B})({\bf C}\otimes{\bf D}) = ({\bf AC})\otimes({\bf BD})$. Letting ${{\bf x}}_t$ denote the $t$th column of ${\bf X} $, the transmitted signal vector on the $M$ transmit antennas at time slot $t$ is given by ${\rm{diag}}( {\bf{c}} )( {{{\bf{1}}_{M/N}} \otimes ({{\bf{Vx}}}_t)} )$ for $t = 1,2,\ldots,T$. Lemma 3 reveals that if and only if all the $N$ elements in ${{\bf Vx}}_t$ have the same amplitude, Requirements 1--2 can be satisfied simultaneously. Therefore, with the precoding matrix in (\ref{General Precoding Matrix}), the remaining work is to design the unitary matrix $\bf V$ therein and the low-dimensional STBC $\bf X$ to let each column ${{\bf Vx}}_t$ of $\bf VX$ have constant amplitude. In Section V, we let $\bf X$ be some existing classic STBCs, e.g., the AC, and then design the unitary matrix $\bf V$ and the constellations for the information symbols in  $\bf X$.

\subsection{Diversity Performance Analysis}

One may treat $\bf WX$ together as an STBC, where $\bf W$ is a precoding matrix independent of information symbols once it was designed, while $\bf X$ includes information symbols. When $\bf WX$ is designed with the proposed approach, with Requirement 1, we know  that it can guarantee equal receiving power at different UTs. However, the diversity performance of this STBC at these UTs is still not clear. In this subsection, we use pairwise error probability (PEP) \cite{V.Tarokh1998,V.Tarokh1999,J.C.Guey1999} to evaluate the diversity performance for our STBC in common information broadcasting.

Consider that there are $K$ UTs in the cell, and the signal model is as (\ref{System Model 2}). For the $k$th UT, the corresponding PEP and its upper bound under maximum likelihood (ML) decoding can be expressed as
\begin{align}
{P_{\mathrm{e},k}} &= \mathbb{E}\{ {\mathbb{P}\{ {{\bf{X}} \to {\bf{X'}}|{{\bf{h}}_k}} \}} \} \nonumber \\
&= \mathbb{E}\Bigg\{ {Q\Bigg( {\sqrt {\frac{{{{\bf{h}}_k^H}{\bf{W}}( {{\bf{X}} - {\bf{X'}}} ){{( {{\bf{X}} - {\bf{X'}}} )^H}}{{\bf{W}}^H}{\bf{h}}_k}}{{2\sigma _{\rm{n}}^2}}} } \Bigg)} \Bigg\} \nonumber \\
& \le \mathbb{E}\Bigg\{ {\exp \Bigg( { - \frac{{{\bf{h}}_k^H{\bf{W}}( {{\bf{X}} - {\bf{X'}}} ){{( {{\bf{X}} - {\bf{X'}}} )^H}}{{\bf{W}}^H}{{\bf{h}}_k}}}{{4\sigma _{\rm{n}}^2}}} \Bigg)} \Bigg\} \nonumber \\
&= \prod\limits_{n = 1}^{r_k} {\frac{1}{{1 + {\lambda _{k,n}}/( {4\sigma _{\rm{n}}^2} )}}} \nonumber \\
&< {(4\sigma _{\rm{n}}^2)}^{r_k} \prod\limits_{n = 1}^{r_k} {{\lambda _{k,n}^{-1}}} \triangleq P_{{\rm{e}},k}^{{\rm{ub}}} \label{PEP Upper Bound Single UT}
\end{align}
where $\bf X \neq {\bf X}' \in \mathcal{C}$ are two distinct codewords in the codebook set $\mathcal{C}$, the expectation is taken over ${\bf h}_k$, the distribution of which is as in (\ref{Channel Model 2}), $Q( x ) = \frac{1}{{\sqrt {2\pi } }} \int_x^\infty  {{e^{ - {t^2}/2}}{\rm{d}}t} $, the first inequality is with $Q( x ) \le {e^{ - {x^2}/2}}$, and $\{\lambda_{k,1},\lambda_{k,2},\ldots,\lambda_{k,r_k}\}$ are the $r_k$ non-zero eigenvalues of
\begin{align}
{\widetilde{\bf R}_k} &\triangleq \mathbb{E}\big\{ {{{( {{\bf{X}} - {\bf{X'}}} )}^H}{{\bf{W}}^H}{{\bf{h}}_k}{\bf{h}}_k^H{\bf{W}}( {{\bf{X}} - {\bf{X'}}} )} \big\} \nonumber \\
&= {( {{\bf{X}} - {\bf{X'}}} )}^H{{\bf{W}}^H}{\bf{F}}_M^H{{\bf{\Lambda }}_k}{{\bf{F}}_M}{\bf{W}}( {{\bf{X}} - {\bf{X'}}} ) , \label{Eigenvalue}
\end{align}
the rank of which is assumed to be $r_k$. In common information broadcasting, it is expected that the transmitted codeword can be decoded successfully by all the UTs. Therefore, the total PEP for all the UTs is defined as the probability for an error-decoding event in at least one of the UTs, i.e.,
\begin{align}
{P_{\rm{e}}} &= 1 - \prod\limits_{k = 1}^K {\Bigg( {1 - \sum_{{\bf X} \neq {\bf X}' \in \mathcal{C}} {P_{{\rm{e}},k}}} \Bigg)} . \label{PEP}
\end{align}
With the upper bound (\ref{PEP Upper Bound Single UT}) for the $k$th UT, the total PEP (\ref{PEP}) can be upper bounded by
\begin{align}
{P_{\rm{e}}} &< 1 - \prod\limits_{k = 1}^K {\Bigg( {1 - \sum_{{\bf X} \neq {\bf X}' \in \mathcal{C}} P_{{\rm{e}},k}^{{\rm{ub}}}} \Bigg)} \nonumber \\
&= 1 - \prod\limits_{k = 1}^K {\Bigg( {1 - \sum_{{\bf X} \neq {\bf X}' \in \mathcal{C}} {{(4\sigma _{\rm{n}}^2)}^{r_k}}\prod\limits_{n = 1}^{r_k} {{\lambda _{k,n}^{-1}}} } \Bigg)}  . \label{PEP Upper Bound 1}
\end{align}
Then we present the following lemma.

\emph{Lemma 4 \cite{X.Meng2016}:} Consider the precoding matrix $\bf W$ in (\ref{General Precoding Matrix}) and the channel covariance matrix ${\bf{F}}_M^H{\bf{\Lambda }}_k{{\bf{F}}_M}$ in (\ref{Channel Model 2}). When the number of BS antennas $M$ goes to infinity and $N$ is kept as a constant, it holds that
\begin{align}
{{\bf{W}}^H}{\bf{F}}_M^H{\bf{\Lambda }}_k{{\bf{F}}_M}{\bf{W}} \xrightarrow{M \to \infty } \frac{1}{N} {{\bf{I}}_N} \nonumber
\end{align}
for any uniformly bounded absolutely integrable function $S_k( \omega  ) = {{2p_k( {\arcsin ( {2\omega} )} )}}/{{\sqrt {1 - 4{\omega ^2}} }}$ over $[-1/2,1/2]$ satisfying $\int_{ 0}^{1} {S_k( \omega  )\mathrm{d}\omega } = 1$, where $p_k( \theta )$ is the channel PAS in (\ref{Channel Model 1}).

With the above lemma and (\ref{Eigenvalue}), we know that in the large-scale array regime, ${\widetilde{\bf R}_1} = {\widetilde{\bf R}_2} = \cdots = {\widetilde{\bf R}_K} = \frac{1}{N}{( {{\bf{X}} - {\bf{X'}}} )}^H( {{\bf{X}} - {\bf{X'}}} )$. Therefore, we have $r_1 = r_2 = \cdots = r_K = r$, ${\lambda _{1,n}} = {\lambda _{2,n}} =  \cdots  = {\lambda _{K,n}} = {\lambda _n}$ for each $n = 1,2,\ldots,r$, while $r$ and $\{ \lambda_{1},\lambda_{2},\ldots,\lambda_{r}\}$ are used to represent the rank and the $r$ non-zero eigenvalues of $\frac{1}{N}{( {{\bf{X}} - {\bf{X'}}} )}^H( {{\bf{X}} - {\bf{X'}}} )$, also the rank and the $r$ non-zero eigenvalues of $\frac{1}{N}( {{\bf{X}} - {\bf{X'}}} ){( {{\bf{X}} - {\bf{X'}}} )}^H$. If ${\bf{X}}$ is an $N\times T$ STBC ($N \leq T$) achieving its full diversity order of $N$, $( {{\bf{X}} - {\bf{X'}}} ){( {{\bf{X}} - {\bf{X'}}} )}^H$ will be with full rank $N$ for any ${{\bf X} \neq {\bf X}' \in \mathcal{C}}$ \cite{V.Tarokh1998}, i.e., $r = N$. In this case, we can express (\ref{PEP Upper Bound 1}) as
\begin{align}
{P_{\rm{e}}} &<  1 - \prod\limits_{k = 1}^K {\Bigg( {1 - \sum_{{\bf X} \neq {\bf X}' \in \mathcal{C}} {{(4\sigma _{\rm{n}}^2)}^{r_k}}\prod\limits_{n = 1}^{r_k} {{\lambda _{k,n}^{-1}}} } \Bigg)} \nonumber \\
&= 1 - {\Bigg( {1 - {{(4\sigma _{\rm{n}}^2)}^N}\sum_{{\bf X} \neq {\bf X}' \in \mathcal{C}}\prod\limits_{n = 1}^N {{\lambda _n^{-1}}} } \Bigg)^K} . \label{PEP Upper Bound 2}
\end{align}
When $\sigma _{\rm{n}}^2 \to 0$, we may assume $0 \le {(4\sigma _{\rm{n}}^2)}^N \cdot \sum_{{\bf X} \neq {\bf X}' \in \mathcal{C}}\prod_{n = 1}^N {\lambda _n^{ - 1}}  \le 1$. Then (\ref{PEP Upper Bound 2}) can be further upper bounded by
\begin{align}
P_{\rm{e}} &< 1 - {\Bigg( {1 - {{(4\sigma _{\rm{n}}^2)}^N}\sum_{{\bf X} \neq {\bf X}' \in \mathcal{C}}\prod\limits_{n = 1}^N {{\lambda _n^{-1}}} } \Bigg)^K} \nonumber \\
&\leq 1 - \Bigg( {1 - K{{(4\sigma _{\rm{n}}^2)}^N} \sum_{{\bf X} \neq {\bf X}' \in \mathcal{C}} \prod\limits_{n = 1}^N {{\lambda _n^{-1}}} } \Bigg) \nonumber \\
&= K{(4\sigma _{\rm{n}}^2)}^N\sum_{{\bf X} \neq {\bf X}' \in \mathcal{C}}\prod\limits_{n = 1}^N {{\lambda _n^{-1}}} \triangleq P_{\rm e}^{\rm ub} , \label{PEP Upper Bound 3}
\end{align}
where the second inequality is with the fact that, for $K \ge 1$,
\begin{align}
{(1 - x)}^K \geq 1 - Kx, \;\; \text{when } 0 \leq x \leq 1. \label{PEP Upper Bound 4}
\end{align}
To verify the correctness of (\ref{PEP Upper Bound 4}), we define a function $f(x) \triangleq {(1 - x)}^K - 1 + Kx$. Since $f(0)=0$ and $f'(x) =  - K{(1 - x)^{K - 1}} + K = K(1 - {(1 - x)^{K - 1}}) \ge 0$ when $0 \leq x \leq 1$, we know that $f(x) \geq 0$ when $0 \leq x \leq 1$. Therefore, we claim (\ref{PEP Upper Bound 4}) holds.

With the upper bound of the total PEP in (\ref{PEP Upper Bound 3}), the diversity order in common information broadcasting is defined as
\begin{align}
d &=  \lim_{\sigma _{\rm{n}}^2 \to 0} \frac{{\log P_{\rm{e}}^{{\rm{ub}}}}}{{\log \sigma _{\rm{n}}^2}} \nonumber \\
&= \mathop {\lim }\limits_{\sigma _{\rm{n}}^2 \to 0} \frac{{\log \big( {K{{(4\sigma _{\rm{n}}^2)}^N}\sum_{{\bf X} \neq {\bf X}' \in \mathcal{C}}\prod\nolimits_{n = 1}^N {{\lambda _n^{-1}}} } \big)}}{{\log \sigma _{\rm{n}}^2}} = N . \label{Diversity Order}
\end{align}
For our STBC design $\bf WX$, the precoding matrix $\bf W$ is channel-independent and cannot provide diversity, hence the diversity is only harvested by the $N\times T$ low-dimensional STBC $\bf X$, which has its maximum diversity order of $N$. Equation (\ref{Diversity Order}) reveals that in the large-scale array regime, the maximum diversity order of $N$ can be achieved by our design. In addition, the diversity order is independent with the number of UTs $K$, since in (\ref{PEP Upper Bound 3}) $K$ does not exist in the exponent term of noise variance $\sigma_\mathrm{n}^2$.

\section{Some STBC Designs}

In this section, we design some detailed examples for the STBC by utilizing the approach proposed in the previous section.

\subsection{Single-Stream Precoding}

We first consider a simple case with $N=T=1$ in (\ref{System Model 2}). Correspondingly, the precoding matrix ${\bf{W}} \in \mathbb{C}^{M \times N}$ and the STBC ${\bf{X}} \in \mathbb{C}^{N \times T}$ in (\ref{System Model 2}) degenerate to a column vector ${\bf{w}} \in \mathbb{C}^{M \times 1}$ and a scalar symbol $x$, respectively. Hence, the transmitted signal at the BS at each time slot is ${\bf w}x$. We have to design the precoding vector $\bf w$ and the constellation of the information symbol $x$ to let the transmitted signal ${\bf w}x$ satisfy Requirements 1--2 in Section III simultaneously, i.e., all the $M$ elements in ${\bf F}_M {\bf w} x$ as well as all the $M$ elements in ${\bf w} x$ have the same amplitude.

Note that the scalar symbol $x$ affects neither Requirement 1 nor 2, i.e., as long as all the $M$ elements in ${\bf F}_M {\bf w} $ as well as all the $M$ elements in ${\bf w} $ have the same amplitude, all the $M$ elements in ${\bf F}_M {\bf w} x$ as well as all the $M$ elements in ${\bf w} x$ will have the same amplitude, for any scalar symbol $x$. With the definition of CAZAC sequences in Section IV, we conclude that $\bf w$ should be a CAZAC sequence of length $M$. Therefore, letting $\bf w$ be a ZC sequence, i.e.,
\begin{align}
{\bf{w}} = {\bf{c}} \label{Precoding Vector}
\end{align}
yields the design for $N = T = 1$, and the constellation of the information symbol $x$ is selected to be phase shift keying (PSK), i.e., $x \in \mathcal{S}_{\mathrm{PSK}} = \{ { 1,e^{j2\pi /L},\ldots,e^{j2\pi ( L-1)/L}  } \}$ for some positive integer $L$, to yield constant instantaneous power at different time slots.

\subsection{Precoded AC}

The above single-stream precoding design just provides spatial diversity order of $1$. To provide spatial diversity order of $2$, we consider the well-known AC \cite{S.M.Alamouti1998}. In this case, we let $N = T = 2$ in (\ref{System Model 2}). Correspondingly, the STBC matrix $\bf X$ in (\ref{System Model 2}) is described as
\begin{align}
{\bf{X}}_{\mathrm{AC}} = {\begin{bmatrix}
{{x_1}}&{x_2^*}\\
{{x_2}}&{ - x_1^*}
\end{bmatrix}} . \label{Alamouti Code}
\end{align}
We have to design the precoding matrix ${\bf{W}}_{\mathrm{AC}} \in {\mathbb{C}^{M \times 2}}$ and the constellations of the two information symbols $x_1$ and $x_2$ in ${\bf{X}}_{\mathrm{AC}}$ to let ${\bf W}_{\mathrm{AC}}{\bf X}_{\mathrm{AC}}   \in {\mathbb{C}^{M \times 2}}$ satisfy Requirements 1--2 simultaneously, i.e., all the $M$ elements in each column of ${\bf F}_M{\bf W}_{\mathrm{AC}}{\bf X}_{\mathrm{AC}}$ have the same amplitude, and all the $M$ elements in each column of ${\bf W}_{\mathrm{AC}}{\bf X}_{\mathrm{AC}} $ also have the same amplitude.

Since $N = 2$, with Lemma 3, we let the number of BS antennas $M$ be an integer multiple of $ 4$. Correspondingly, the precoding matrix is proposed to be
\begin{align}
{\bf{W}}_{\mathrm{AC}} = {\rm{diag}}( {\bf{c}} )( {{{\bf{1}}_{M/2}} \otimes {{\bf{I}}_2}} ) . \label{Precoded Alamouti Code Precoding Matrix}
\end{align}
Therefore, the transmitted signal at the BS is ${\bf W}_{\mathrm{AC}}{\bf X}_{\mathrm{AC}}  = {\rm{diag}}( {\bf{c}} )( {{\bf{1}}_{M/2}} \otimes {\bf X}_{\mathrm{AC}}  )$. Under this structure, Lemma 3 reveals that if and only if the two information symbols $x_1$ and $x_2$ in (\ref{Alamouti Code}) have the same amplitude, Requirements 1--2 can be satisfied simultaneously. To guarantee that $x_1$ and $x_2$ are with the same amplitude for any realization in their constellation sets, the constellations of both $x_1$ and $x_2$ are selected to be PSK, i.e., ${{x_1},{x_2}} \in {\mathcal{S}}_{\mathrm{PSK}}$.

It is noted that, in our previous paper \cite{X.Meng2014}, the precoding matrix ${\bf W}_{\mathrm{AC}}$ is proposed to be ${\rm{diag}}( {\bf{c}} )( {{\bf{1}}_{M/2}}  \otimes {\bf{H}}_2 )$ where ${\bf{H}}_2 $ is the $2 \times 2$ Hadamard matrix. It is not hard to see that the PSK constellations of $x_1$ and $x_2$ used in this paper has the same minimum Euclidean distance with that of the joint design proposed in \cite{X.Meng2014}. However, since $x_1$ and $x_2$ are modulated independently in this paper while they are modulated jointly in \cite{X.Meng2014}, they can be decoded separately in this paper while they must be decoded jointly in \cite{X.Meng2014}. Therefore, the proposed design in this paper has the same coding gain with the joint design in \cite{X.Meng2014} while the decoding complexity can be reduced.

\subsection{Precoded OSTBC}

Since the maximum achievable diversity order of AC is only $2$, if a higher diversity order is desired, we need to consider the STBC $\bf{X}$ in (\ref{System Model 2}) with a larger size \cite{V.Tarokh1999}. One option is to use orthogonal STBCs (OSTBCs) \cite{V.Tarokh1999,X.B.Liang2003,W.Su2003,K.Lu2005,G.Ganesan2001,O.Tirkkonen2002,B.M.Hochwald2001}. These codes, including the $2 \times 2$ AC as a special case, achieve full diversity and have symbol-wise ML decoding at the receiver side. Here, we consider the following well-known OSTBC with symbol rate of $3/4$, \cite{G.Ganesan2001,O.Tirkkonen2002,B.M.Hochwald2001},
\begin{align}
{\bf{X}}_{\mathrm{OS}} = {\begin{bmatrix}
{{x_1}}&{x_2^*}&{x_3^*}&0\\
{{x_2}}&{ - x_1^*}&{\rm{0}}&{x_3^*}\\
{{x_3}}&{\rm{0}}&{ - x_1^*}&{ - x_2^*}\\
{\rm{0}}&{{x_3}}&{ - {x_2}}&{{x_1}}
\end{bmatrix}} . \label{OSTBC}
\end{align}
We have to design the precoding matrix ${\bf{W}}_{\mathrm{OS}} \in {\mathbb{C}^{M \times 4}}$ and the constellations of the three information symbols $x_1,x_2,x_3$ in ${\bf{X}}_{\mathrm{OS}}$ to let ${\bf W}_{\mathrm{OS}} {\bf X}_{\mathrm{OS}}  \in {\mathbb{C}^{M \times 4}}$ satisfy Requirements 1--2 simultaneously, i.e., all the $M$ elements in each column of ${\bf F}_M{\bf W}_{\mathrm{OS}} {\bf X}_{\mathrm{OS}}$ as well as all the $M$ elements in each column of ${\bf W}_{\mathrm{OS}} {\bf X}_{\mathrm{OS}}$ have the same amplitude.

From (\ref{OSTBC}), one can see that there are zero entries in the codeword matrix ${\bf X}_{\mathrm{OS}}$, while others are information symbols $x_n$. This means that the signal power on the four virtual transmit ports using ${\bf X}_{\mathrm{OS}}$ in (\ref{OSTBC}) will be different no matter how one designs a constellation for $x_n$. Therefore the identity precoding matrix ${\bf I}_4$ may not be used directly as what is done with ${\bf I}_2$ in (\ref{Precoded Alamouti Code Precoding Matrix}) for the AC. As an alternative method, the precoding matrix here is proposed to be
\begin{align}
{\bf{W}}_{\mathrm{OS}} = {\rm{diag}}( {\bf{c}} )( {{{\bf{1}}_{M/4}} \otimes {{{\bf{I}}_2} \otimes {{\bf{H}}_2}} } ) \label{Precoded OSTBC Precoding Matrix}
\end{align}
where $\bf{c}$ is defined in (\ref{ZC Sequence}) and
\begin{align}
{{\bf{H}}_2} = \frac{1}{{\sqrt 2 }}{\begin{bmatrix}
1&1\\
1&{ - 1}
\end{bmatrix}}  \label{Hadamard Matrix}
\end{align}
is the $2 \times 2$ unitary Hadamard matrix. One can see that, multiplying the OSTBC matrix ${\bf X}_{\mathrm{OS}}$ in (\ref{OSTBC}) with the precoding
matrix ${\bf W}_{\mathrm{OS}}$ in (\ref{Precoded OSTBC Precoding Matrix}) is just to repeat
\begin{align}
( {{{\bf{I}}_2} \otimes {{\bf{H}}_2}} ){\bf{X}}_{\mathrm{OS}} = \frac{1}{{\sqrt 2 }}{\begin{bmatrix}
{{x_1} + {x_2}}&{x_2^* - x_1^*}&{x_3^*}&{x_3^*}\\
{{x_1} - {x_2}}&{x_2^* + x_1^*}&{x_3^*}&{ - x_3^*}\\
{{x_3}}&{{x_3}}&{ - x_1^* - {x_2}}&{{x_1} - x_2^*}\\
{{x_3}}&{ - {x_3}}&{ - x_1^* + {x_2}}&{ - {x_1} - x_2^*}
\end{bmatrix}} \label{OSTBC After Precoding}
\end{align}
periodically across transmit antennas, and then adjust the phases via ZC sequence $\bf c$ on each antenna. The difference between the codeword matrix (\ref{OSTBC After Precoding}) after precoding and the codeword matrix (\ref{OSTBC}) before precoding is that, there is no zero entry in (\ref{OSTBC After Precoding}) any more and this leads to the possibility to have constant power across transmit antennas by designing the constellations for information symbols $x_n$ properly. Let the number of BS antennas $M$ be an integer multiple of $16$. With Lemma 3, we know that Requirements 1--2 can be satisfied simultaneously if and only if all the four elements in each column of $( {{{\bf{I}}_2} \otimes {{\bf{H}}_2}} ) {\bf{X}}_{\mathrm{OS}}$ in (\ref{OSTBC After Precoding}) have the same amplitude. Readily, we conclude that the three information symbols $x_1,x_2,x_3$ in (\ref{OSTBC After Precoding}) should satisfy
\begin{align}
| {{x_1} + {x_2}} | = | {{x_1} - {x_2}} | = | {x_1^* + {x_2}} | = | {x_1^* - {x_2}} | = | {{x_3}} | . \label{Precoded OSTBC Constellation Constraint}
\end{align}

We start the constellation design for $x_1$ and $x_2$, while $x_3$ will be considered latter. It is not hard to see that, to let (\ref{Precoded OSTBC Constellation Constraint}) hold, $x_1$ must be orthogonal to both $x_2$ and $x_2^*$ simultaneously, if we consider $x_1$ and $x_2$ as two $2$-dimensional real vectors in the complex plane. This condition can be satisfied if and only if one of $x_1$ and $x_2$ is on the real axis and the other is on the imaginary axis. Therefore, the constellations of $x_1$ and $x_2$ are selected to be pulse amplitude modulation (PAM), i.e., ${x_1} \in \mathcal{S}_{\mathrm{PAM}}$ and ${x_2} \in j \mathcal{S}_{\mathrm{PAM}}$, where $\mathcal{S}_{\mathrm{PAM}} = d \cdot \{ {\pm 1, \pm 3, \ldots , \pm ( {2L - 1} )} \}$ while $L$ determines the modulation order and $d$ normalizes the average power, respectively. Then we consider $x_3$. Obviously, once $x_1$ and $x_2$ are obtained after modulation symbol mapping, $|x_3|$ should be equal to $| {{x_1} + {x_2}} |$ to let (\ref{Precoded OSTBC Constellation Constraint}) hold. Hence, the amplitude of $x_3$ has no degree-of-freedom and cannot be modulated with information, and only the phase of $x_3$ can be modulated with information. Consequently, the constellation of $x_3$ is selected to be PSK with varied amplitude, i.e., ${x_3} \in | {{x_1} + {x_2}} | \cdot \mathcal{S}_{\mathrm{PSK}}$. Moreover, it is also expected that the constellations of different symbols have the same minimum Euclidean distance. Otherwise, the overall performance will be dominated by the ``weakest'' signal constellation that has the smallest minimum Euclidean distance. To this end, we have the following design. Let $R$ denote the bit rate of (\ref{OSTBC}), i.e., $4R$ bits are transmitted within four time slots. For these $4R$ bits, $2R-1$ bits are mapped to $x_1$ where $x_1 \in \mathcal{S}_{m\mathrm{PAM}}$ and $m = 2^{2R-1}$, $2R-1$ bits are mapped to $x_2$ where $x_2 \in j\mathcal{S}_{m\mathrm{PAM}}$, and $2$ bits are mapped to $x_3$ where ${x_3} \in | {{x_1} + {x_2}} | \cdot \{ \pm1, \pm j\} $. It can be verified that this design guarantees the same minimum Euclidean distance for all the three constellations of $x_1,x_2,x_3$.

Furthermore, we consider the ML decoding procedure for the above encoding. Note that the codeword matrix defined in (\ref{OSTBC}) with the above constellation design can also be equivalent to the following codeword matrix
\begin{align}
{\bf{X}}'_{\mathrm{OS}} = {\begin{bmatrix}
{{x_1}}&{x_2^*}&{| {{x_1} + {x_2}} |{(x_3')}^*}&0\\
{{x_2}}&{ - x_1^*}&0&{| {{x_1} + {x_2}} |{(x'_3)}^*}\\
{| {{x_1} + {x_2}} |{x'_3}}&0&{ - x_1^*}&{ - x_2^*}\\
0&{| {{x_1} + {x_2}} |{x'_3}}&{ - {x_2}}&{{x_1}}
\end{bmatrix}}  \label{OSTBC Revised}
\end{align}
where $x_1 \in \mathcal{S}_{m\mathrm{PAM}}$, $x_2 \in j \mathcal{S}_{m\mathrm{PAM}}$, and $x'_3 \in \mathcal{S}_{\mathrm{QPSK}}$. In this case, $x'_3$ will be independent with $x_1$ and $x_2$. With (\ref{System Model 2}) and (\ref{OSTBC Revised}), and omit the UT index $k$, the logarithm likelihood function of $x_1,x_2,x'_3$ can be expressed as
\begin{align}
\ell ( {{x_1},{x_2},{x'_3}} ) & \propto ( {{{| {{x_1}} |}^2} + {{| {{x_2}} |}^2}} )\sum\limits_{n = 1}^4 {{{| {{g_n}} |}^2}}  - 2| {{x_1} + {x_2}} |\underbrace {\Re ( {{x'_3}( {{g_3}y_1^* + {g_4}y_2^*} ) + {(x'_3)}^*( {{g_1}y_3^* + {g_2}y_4^*} )} )}_{f( {{x'_3}} )} \nonumber \\
& \;\;\;\; + {| {{x_1}} |}^2\sum\limits_{n = 1}^4 {{{| {{g_n}} |}^2}}  - 2\Re ( {{x_1}( {{g_1}y_1^* + {g_4}y_4^*} ) - x_1^*( {{g_2}y_2^* + {g_3}y_3^*} )} ) \nonumber \\
& \;\;\;\; + {| {{x_2}} |}^2\sum\limits_{n = 1}^4 {{{| {{g_n}} |}^2}}  - 2\Re ( {{x_2}( {{g_2}y_1^* - {g_4}y_3^*} ) + x_2^*( {{g_1}y_2^* - {g_3}y_4^*} )} ) \label{Precoded OSTBC ML Decoding}
\end{align}
where $[ {{g_1}, {g_2}, g_3 ,{g_4}} ] \triangleq {{\bf{h}}^H}{\bf{W}}_{\mathrm{OS}}$. From (\ref{Precoded OSTBC ML Decoding}), one can see that the minimization of $\ell ( {{x_1},{x_2},{x'_3}} )$ with respect to $x_1$, $x_2$, and $x'_3$ in their constellation sets for ML decoding can be divided into two steps. First, one can maximize $f( x'_3 )$ with respect to $x'_3$. This means that $x'_3$ can be decoded independently. With the decoded value of $x'_3$, assumed to be $\widehat x'_3$, one can then minimize $\ell( {{x_1},{x_2},{\hat x'_3}} )$ with respect to $x_1$ and $x_2$. However, $x_1$ and $x_2$ must be decoded jointly since they cannot be separated from each other owing to the term $| {{x_1} + {x_2}} |$ in (\ref{Precoded OSTBC ML Decoding}), albeit that they are modulated independently. This means that the proposed precoded OSTBC does not have symbol-wise ML decoding any more, but has pair-wise ML decoding.

\subsection{Precoded QOSTBC}

Although the above OSTBC can provide diversity order of $4$, its symbol rate is only $3/4$, which is also the upper bound of the symbol rate of any OSTBC for more than two transmit antennas \cite{H.Q.Wang2003}. To achieve a higher symbol rate, one approach is to relax the orthogonality, i.e., quasi-orthogonal STBC (QOSTBC) \cite{O.Tirkkonen2000,H.Jafarkhani2001}. With the quasi-orthogonal structure, the ML decoding at the receiver can be done by searching pairs of symbols. Moreover, the signal constellations can be properly designed to achieve full diversity with maximized coding gain \cite{W.Su2004}.

Here, we consider the case with $N=T=4$ in (\ref{System Model 2}). Correspondingly, the STBC matrix $\bf X$ in (\ref{System Model 2}) is described as the Tirkkonen, Boariu, and Hottinen (TBH) quasi-orthogonal STBC (QOSTBC) matrix \cite{H.Jafarkhani2001,O.Tirkkonen2000}
\begin{align}
{\bf{X}}_{\mathrm{QO}} = {\begin{bmatrix}
{{x_1}}&{x_2^*}&{{x_3}}&{x_4^*}\\
{{x_2}}&{ - x_1^*}&{{x_4}}&{ - x_3^*}\\
{{x_3}}&{x_4^*}&{{x_1}}&{x_2^*}\\
{{x_4}}&{ - x_3^*}&{{x_2}}&{ - x_1^*}
\end{bmatrix}} . \label{QOSTBC}
\end{align}
We have to design the precoding matrix ${\bf{W}}_{\mathrm{QO}} \in {\mathbb{C}^{M \times 4}}$ and the constellations of the four information symbols $x_1,x_2,x_3,x_4$ in ${\bf{X}}_{\mathrm{QO}}$ to let ${\bf W}_{\mathrm{QO}} {\bf X}_{\mathrm{QO}}  \in {\mathbb{C}^{M \times 4}}$ satisfy Requirements 1--2 simultaneously, i.e., all the $M$ elements in each column of ${\bf F}_M{\bf W}_{\mathrm{QO}} {\bf X}_{\mathrm{QO}}$ as well as all the $M$ elements in each column of ${\bf W}_{\mathrm{QO}} {\bf X}_{\mathrm{QO}}$ have the same amplitude.

Since $N = 4$, with Lemma 3, we let the number of BS antennas $M$ be an integer multiple of $16$. Correspondingly, the precoding matrix is proposed to be
\begin{align}
{\bf W}_{\mathrm{QO}} = {\rm{diag}}( {\bf{c}} )( {{{\bf{1}}_{M/4}} \otimes {{\bf{I}}_4}} ) . \label{Precoded QOSTBC Precoding Matrix}
\end{align}
Hence, the transmitted signal is ${\bf W}_{\mathrm{QO}} {\bf X}_{\mathrm{QO}} = {\rm{diag}}( {\bf{c}} )( {{\bf{1}}_{M/4}} \otimes  {\bf X}_{\mathrm{QO}} )$. Lemma 3 tells us that Requirements 1--2 can be satisfied simultaneously if and only if the four information symbols $x_1,x_2,x_3,x_4$ in (\ref{QOSTBC}) have the same amplitude. Therefore, all the constellations of $x_1,x_2,x_3,x_4$ are selected to be PSK. Moreover, to guarantee that diversity order of $4$ can be achieved, a typical method is to let ${{x_1},{x_3}} \in {\mathcal{S}}_{\mathrm{PSK}}$ and ${{x_2},{x_4}} \in { e^{j\Theta}\mathcal{S}_{\mathrm{PSK}}}$. The optimal rotation angle to maximize the coding gain is proven to be $\Theta = \pi/L$ when $L$ is even and $\Theta = \pi/(2L)$ when $L$ is odd, respectively \cite{D.Wang2005}. Furthermore, although $x_1,x_2,x_3,x_4$ are modulated independently, for ML decoding, $x_1$ and $x_3$ need to be decoded jointly, while $x_2$ and $x_4$ need to be decoded jointly \cite{W.Su2004}.

\subsection{Precoded CIOD}

The precoding matrix in (\ref{Precoded QOSTBC Precoding Matrix}) and the PSK constellation for the four information symbols in (\ref{QOSTBC}) can be seen as a trivial generalization of the precoding matrix in (\ref{Precoded Alamouti Code Precoding Matrix}) and the PSK constellation for the two information symbols in (\ref{Alamouti Code}). One will ask whether there exists another design to outperform precoded QOSTBC? To answer this question, we consider the coordinate interleaved orthogonal design (CIOD) first proposed by Khan-Rajan \cite{Z.A.Khan2002,Z.A.Khan2006}, where they placed an OSTBC on diagonal repeatedly with different information symbols and then these different information symbols are interleaved in such a way that the final overall design has full diversity.

The same with precoded QOSTBC, we let $N=T=4$ in (\ref{System Model 2}). Correspondingly, the STBC matrix $\bf X$ in (\ref{System Model 2}) is described as the CIOD matrix
\begin{align}
{\bf{X}}_{\mathrm{CI}} = {\begin{bmatrix}
{{x_1}}&{x_2^*}&0&0\\
{{x_2}}&{ - x_1^*}&0&0\\
0&0&{{x_{\rm{3}}}}&{x_4^*}\\
0&0&{{x_{\rm{4}}}}&{ - x_{\rm{3}}^*}
\end{bmatrix}} \label{CIOD}
\end{align}
where the four information symbols $x_1,x_2,x_3,x_4$ should be interleaved in a specific pattern to achieve diversity order of $4$ \cite{Z.A.Khan2002,Z.A.Khan2006}. We have to design the precoding matrix ${\bf{W}}_{\mathrm{CI}} \in {\mathbb{C}^{M \times 4}}$ as well as the constellations and the interleaving pattern of the four information symbols in ${\bf{X}}_{\mathrm{CI}}$ to let ${\bf W}_{\mathrm{CI}} {\bf X}_{\mathrm{CI}}  \in {\mathbb{C}^{M \times 4}}$ satisfy Requirements 1--2 simultaneously, i.e., all the $M$ elements in each column of ${\bf F}_M{\bf W}_{\mathrm{CI}} {\bf X}_{\mathrm{CI}} $ have the same amplitude, and all the $M$ elements in each column of ${\bf W}_{\mathrm{CI}} {\bf X}_{\mathrm{CI}} $ also have the same amplitude.

From (\ref{CIOD}), one can see that there are zero entries in each column of the codeword matrix ${\bf X}_{\mathrm{CI}} $, while others are information symbols $x_n$. This means that at any time slot, the four virtual transmit ports using ${\bf X}_{\mathrm{CI}}$ always have different signal power no matter how one designs a constellation for $x_n$ therein. Therefore, the identity matrix ${\bf I}_4$ may not be used directly for the precoding matrix ${\bf W}_{\mathrm{CI}}$ as what is done for the precoding matrix ${\bf W}_{\mathrm{QO}}$ in (\ref{Precoded QOSTBC Precoding Matrix}). Hence, we propose to use ${{{\bf{H}}_2} \otimes {{\bf{H}}_2}}$ instead of ${\bf I}_4$, as an alternative realization of the unitary matrix $\bf V$ in (\ref{General Precoding Matrix}), where ${{\bf{H}}_2}$ is the $2 \times 2$ unitary Hadamard matrix as (\ref{Hadamard Matrix}). Correspondingly, the precoding matrix ${\bf W}_{\mathrm{CI}} $ is proposed to be
\begin{align}
{\bf{W}}_{\mathrm{CI}} = {\rm{diag}}( {\bf{c}} )( {{{\bf{1}}_{M/4}} \otimes {{{\bf{H}}_2} \otimes {{\bf{H}}_2}}} ) \label{Precoded CIOD Precoding Matrix}
\end{align}
where $M$ needs to be an integer multiple of $16$. Therefore, the transmitted signal is ${\bf{W}}_{\mathrm{CI}}{\bf{X}}_{\mathrm{CI}} = {\rm{diag}}( {\bf{c}} )( {{{\bf{1}}_{M/4}} \otimes (({{{\bf{H}}_2} \otimes {{\bf{H}}_2}}) {{\bf{X}}_{\mathrm{CI}}} )} )$ where
\begin{align}
( {{{\bf{H}}_2} \otimes {{\bf{H}}_2}} ){\bf{X}}_{\mathrm{CI}} = \frac{1}{2} {\begin{bmatrix}
{{x_1} + {x_2}}&{x_2^* - x_1^*}&{{x_{\rm{3}}} + {x_{\rm{4}}}}&{x_4^* - x_{\rm{3}}^*}\\
{{x_1} - {x_2}}&{x_1^* + x_2^*}&{{x_{\rm{3}}} - {x_{\rm{4}}}}&{x_{\rm{3}}^* + x_4^*}\\
{{x_1} + {x_2}}&{x_2^* - x_1^*}&{ - {x_{\rm{3}}} - {x_{\rm{4}}}}&{x_{\rm{3}}^* - x_4^*}\\
{{x_1} - {x_2}}&{x_1^* + x_2^*}&{{x_{\rm{4}}} - {x_{\rm{3}}}}&{ - x_{\rm{3}}^* - x_4^*}
\end{bmatrix}} . \label{CIOD After Precoding}
\end{align}
The difference between the codeword matrix after precoding in (\ref{CIOD After Precoding}) and that before precoding in (\ref{CIOD}) is that, due to the interleaving effect of ${{{\bf{H}}_2} \otimes {{\bf{H}}_2}}$, there is no zero entry in (\ref{CIOD After Precoding}) any more and this leads to the possibility to have constant power across transmit antennas by designing the constellations for information symbols $x_n$ properly. For the transmitted signal ${\bf{W}}_{\mathrm{CI}}{\bf{X}}_{\mathrm{CI}}$, Lemma 3 tells us that Requirements 1--2 can be satisfied simultaneously if and only if $x_1+x_2$ and $x_1-x_2$ have the same amplitude, and $x_3+x_4$ and $x_3-x_4$ also have the same amplitude, i.e.,
\begin{align}
|x_1+x_2| &= |x_1-x_2|   \label{CIOD Modulus 1} \\
|x_3+x_4| &= |x_3-x_4| . \label{CIOD Modulus 2}
\end{align}
Then, we propose the following encoding procedure.
\begin{itemize}
\item First, map binary bits to two information symbols $s_1,s_2 \in e^{j\Theta}{\mathcal{S}}_{\mathrm{QAM}}$, where $\mathcal{S}_{\mathrm{QAM}} = d \cdot \{ \pm 1 \pm j, \pm 3 \pm j3, \ldots , \pm ( 2L - 1 )\pm j( 2L - 1 ) \}$ is the quadrature amplitude modulation (QAM) constellation, $d$ is used to normalize the average power of the constellation points, and $\Theta = \arctan ( 2 )/2$ is a pre-defined rotation angle;
\item Then, define the interleaved symbols
\begin{align}
\left\{ {\begin{matrix*}[l]
{{x_1} = \sqrt{2} ( {1 + j} )\Re ( {{s_1}} )}\\
{{x_2} = \sqrt{2} ( {1 - j} )\Re ( {{s_2}} )}\\
{{x_3} = \sqrt{2} ( {1 + j} )\Im ( {{s_1}} )}\\
{{x_4} = \sqrt{2} ( {j - 1} )\Im ( {{s_2}} )} ;
\end{matrix*}} \right. \label{CIOD Interleaving}
\end{align}
\item At last, transmit the precoded CIOD matrix ${\bf{W}}_{\mathrm{CI}}{\bf{X}}_{\mathrm{CI}}$ with (\ref{CIOD}) and (\ref{Precoded CIOD Precoding Matrix}).
\end{itemize}
With the interleaving pattern in (\ref{CIOD Interleaving}), it can be verified that (\ref{CIOD Modulus 1}) and (\ref{CIOD Modulus 2}) can be satisfied, i.e., Requirements 1--2 can be satisfied. Moreover, the rotation angle $\Theta = \arctan ( 2 )/2$ being used to generate $s_1$ and $s_2$ guarantees diversity order of $4$ for (\ref{CIOD}) and is optimal to maximize the coding gain \cite{Z.A.Khan2002,Z.A.Khan2006,H.Q.Wang2009}.

The main difference between the encoding procedure in our CIOD and that in previous CIODs, e.g., \cite{Z.A.Khan2002,Z.A.Khan2006,H.Q.Wang2009}, is that, only two actual information symbols $s_1$ and $s_2$ are transmitted in our design while four symbols can be transmitted in previous designs, i.e., the actual symbol rate is $1/2$ for our code while it is $1$ for previous codes. This reduction of symbol rate can be seen as the price to satisfy Requirements 1--2. Similar price also exists in precoded QOSTBC designed in the previous subsection, where the price is not on the symbol rate, but on the signal constellation, since the signal constellation for QOSTBC is confined to PSK while more energy-efficient and commonly used QAM cannot be applied there.

\emph{Comparison between precoded OSTBC, precoded QOSTBC, and precoded CIOD:} From signal model (\ref{System Model 2}), we can seen ${\bf W}^H{\bf h}$ as an effective channel, and the low-dimensional STBC $\bf X$ is just transmitted over ${\bf W}^H{\bf h}$. Lemma 4 tells us that for all of the three precoding matrices ${\bf W}_{\mathrm{OS}}$, ${\bf W}_{\mathrm{QO}}$, and ${\bf W}_{\mathrm{CI}}$ in (\ref{Precoded OSTBC Precoding Matrix}), (\ref{Precoded QOSTBC Precoding Matrix}), and (\ref{Precoded CIOD Precoding Matrix}), the corresponding three effective channel vectors approach i.i.d. Rayleigh fading, i.e., ${\bf W}^H_{\mathrm{OS}}{\bf h} \sim \mathcal{CN}({\bf 0},\frac{1}{4}{\bf I}_4)$, ${\bf W}^H_{\mathrm{QO}}{\bf h} \sim \mathcal{CN}({\bf 0},\frac{1}{4}{\bf I}_4)$, and ${\bf W}^H_{\mathrm{CI}}{\bf h} \sim \mathcal{CN}({\bf 0},\frac{1}{4}{\bf I}_4)$, when the number of BS antennas $M$ is sufficiently large. Therefore, we can compare the performance among the three low-dimensional STBCs ${\bf{X}}_{\mathrm{OS}}$, ${\bf{X}}_{\mathrm{QO}}$, and ${\bf{X}}_{\mathrm{CI}}$ under the i.i.d. Rayleigh fading channel, instead of comparing the three high-dimensional STBCs ${\bf{W}}_{\mathrm{OS}}{\bf{X}}_{\mathrm{OS}}$, ${\bf{W}}_{\mathrm{QO}}{\bf{X}}_{\mathrm{QO}}$, and ${\bf{W}}_{\mathrm{CI}}{\bf{X}}_{\mathrm{CI}}$ directly.

Coding gain, also known as diversity product in some literature, is a commonly used criterion to evaluate the performance of an STBC with ML decoding at the receiver side. It is desired that the coding gain, which is defined as
\begin{align}
\xi &= \mathop {\min }\limits_{{\bf{X}} \ne {\bf{X'}} \in \mathcal{C} } {\Big( {\det \Big( {( {{\bf{X}} - {\bf{X'}}} ){{( {{\bf{X}} - {\bf{X'}}} )}^H}} \Big)} \Big)^{1/T}} \nonumber
\end{align}
where $\bf X$ and $\bf X'$ are two distinct codewords in the codebook set $\mathcal{C} $, is as large as possible \cite{W.Su2004}.

Owing to the orthogonal property of the OSTBC in (\ref{OSTBC}), i.e.,
\begin{align}
{\bf{X}}_{\mathrm{OS}}{{\bf{X}}_{\mathrm{OS}}^H} = ( {{{| {{x_1}} |}^2} + {{| {{x_2}} |}^2} + {{| {{x_3}} |}^2}} ){{\bf{I}}_4}, \nonumber
\end{align}
the coding gain is the minimum squared Euclidean distance of the signal constellation. An example is shown how to calculate the minimum squared Euclidean distance for different bit rates with our proposed constellation design. Consider the bit rate with $2$ bps/Hz, i.e., totally $8$ bits are transmitted within four time slots in (\ref{OSTBC}). As mentioned in the constellation design for (\ref{OSTBC After Precoding}), in this case, $3$ bits are mapped to ${x_1} \in \mathcal{S}_{8\mathrm{PAM}} $, $3$ bits are mapped to ${x_2} \in j\mathcal{S}_{8\mathrm{PAM}} $, and $2$ bits are mapped to ${x_3} \in | {{x_1} + {x_2}} | \cdot \mathcal{S}_{\mathrm{QPSK}} $. We can obtain that the minimum squared Euclidean distance of the constellations is $4/21$, which is also the minimum squared Euclidean distance of 8PAM. For the QOSTBC matrix ${\bf X}_{\mathrm{QO}}$ in (\ref{QOSTBC}) with PSK constellation, its coding gain has been shown in \cite{D.Wang2005} as
\begin{align}
\xi_{\mathrm{QO}} = \left\{ {\begin{matrix*}[l]
{4{{\sin }^2}( {\pi /L} ),}&{L \le 6}\\
{8{{\sin }^3}( {\pi /L} ),}&{L > 6}
\end{matrix*}} \right. \nonumber
\end{align}
where $\log_2 L$ denotes the bit rate. As an example, when the bit rate is $1$ bps/Hz, i.e., $L = 2$, we can obtain $\xi_{\mathrm{QO}} = 4$. The coding gain of the CIOD matrix ${\bf X}_{\mathrm{CI}}$ in (\ref{CIOD}) with the proposed constellation and interleaving pattern can be expressed as
\begin{align}
\xi_{\mathrm{CI}} &= \mathop {\min }\limits_{{\bf{X}}_{\mathrm{CI}} \ne {\bf{X}}'_{\mathrm{CI}} \in \mathcal{C} } {\Big( {\det \Big( {( {{\bf{X}}_{\mathrm{CI}} - {\bf{X}}'_{\mathrm{CI}}} ){{( {{\bf{X}}_{\mathrm{CI}} - {\bf{X}}'_{\mathrm{CI}}} )}^H}} \Big)} \Big)^{1/4}} \nonumber \\
&= \mathop {\min }\limits_{{s_1} \ne {{s}_1'}\in {e^{j\Theta}}\mathcal{S}_{\mathrm{QAM}}} 4 \cdot | {\Re ( {{s_1} - {s_1'}} )} {\Im ( {{s_1} - {{s}_1'}} )} | \nonumber \\
&= 16{d^{2}}\cos \Theta \sin \Theta \nonumber \\
& = \frac{{16{d^2}}}{{\sqrt 5 }} \label{CIOD Diversity Product}
\end{align}
where the second equation is with (\ref{CIOD}) and (\ref{CIOD Interleaving}), the last equation is with $\sin ( {2\Theta } ) = 2/\sqrt 5 $ since $\Theta  = \arctan ( 2 )/2$, and $d$ denotes the minimum constellation distance for QAM. One can use (\ref{CIOD Diversity Product}) to calculate the coding gains for different bit rates. For example, when the bit rate is $1$ bps/Hz, i.e., $4$ bits are transmitted within the four time slots in (\ref{CIOD}), each one of $s_1$ and $s_2$ in (\ref{CIOD Interleaving}) needs to carry $2$ bits since they are the actual symbols carrying information. In this case, $\mathcal{S}_{\mathrm{QAM}}$ in (\ref{CIOD Interleaving}) corresponds to QPSK, i.e, $d = 1/\sqrt{2}$. Then we can calculate $\xi_{\mathrm{CI}} = 8/\sqrt{5}$ with (\ref{CIOD Diversity Product}).

\begin{figure}[t]
\centering
\includegraphics[width=0.75\columnwidth]{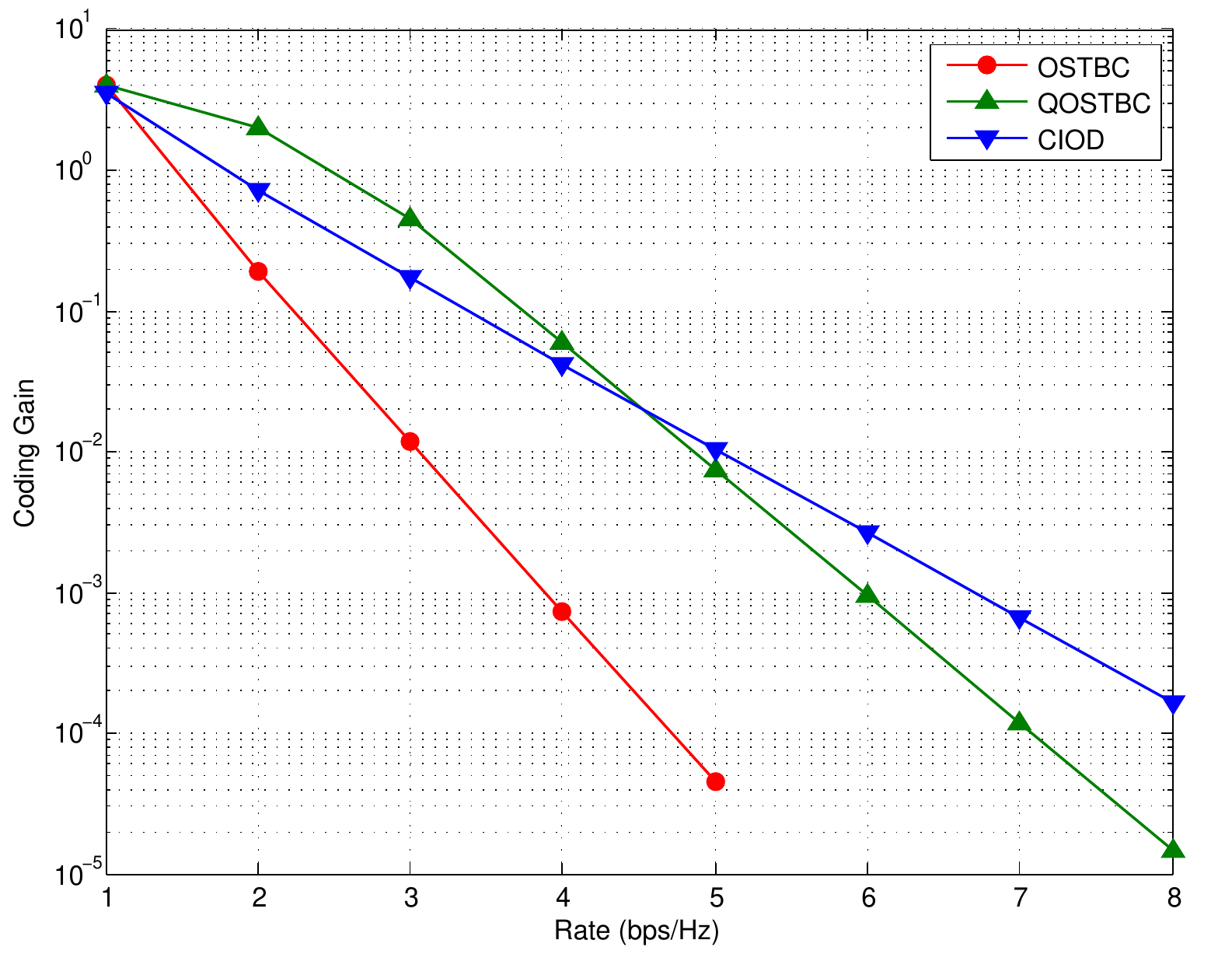}
\caption{Coding gain comparison between different designs for different bit rates.}
\end{figure}

\begin{figure}[t]
\centering
\includegraphics[width=0.75\columnwidth]{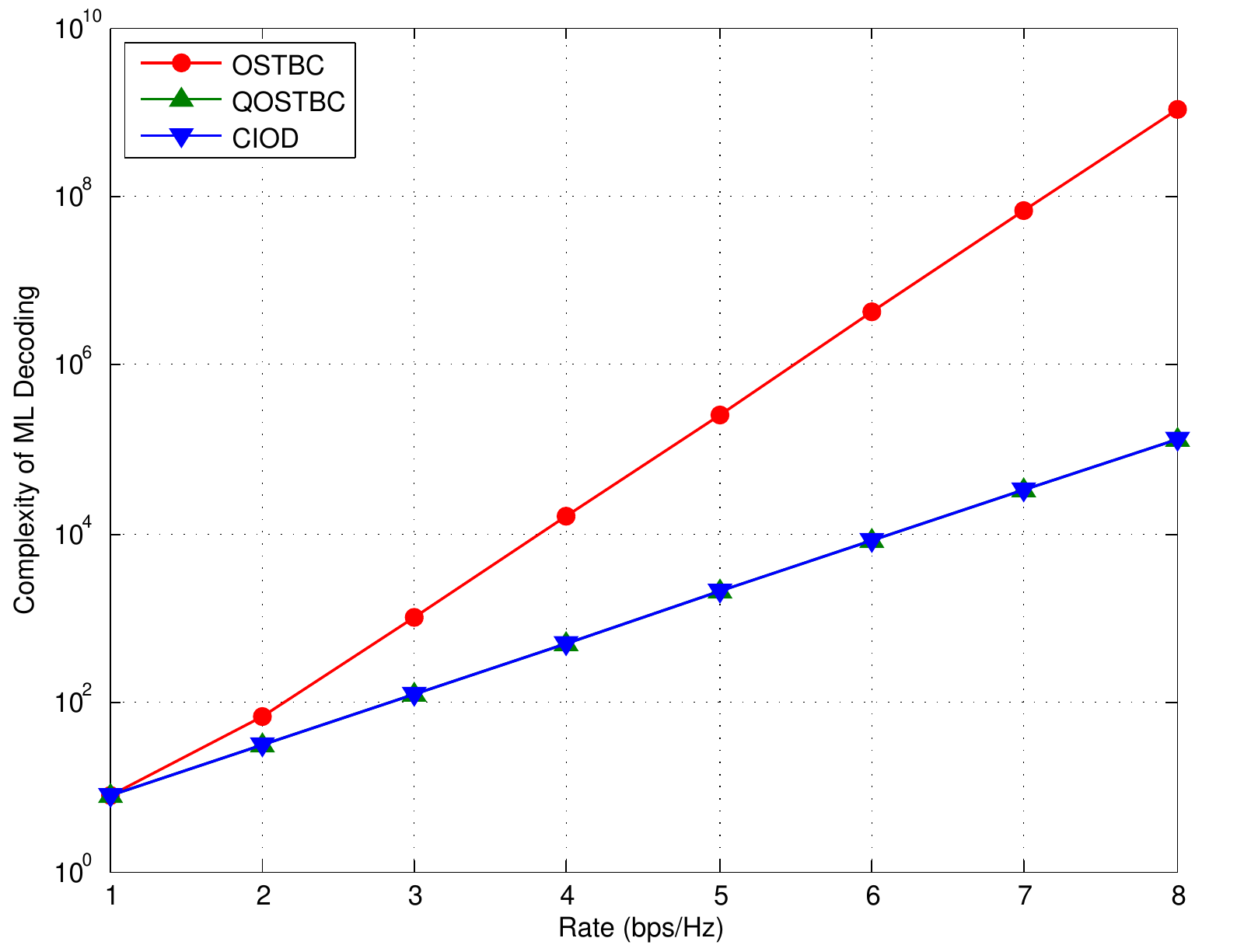}
\caption{Complexity of ML decoding comparison between different designs for different bit rates.}
\end{figure}

The coding gains of OSTBC, QOSTBC, and CIOD designed in this paper for different bit rates are compared in Fig. 1. It is observed that, QOSTBC has a higher coding gain when the bit rate is lower than or equal to $4$ bps/Hz, while CIOD has a higher coding gain when the bit rate is higher than $4$ bps/Hz, and this gap deepens when the bit rate goes high. Therefore, QOSTBC is more favorable when the bit rate is lower than or equal to $4$ bps/Hz while CIOD is more favorable when the bit rate is higher than $4$ bps/Hz. The coding gain of OSTBC is the same with that of QOSTBC at $1$ bps/Hz, but is much lower than QOSTBC and CIOD when the bit rate goes high. This is because higher order modulation is needed in OSTBC to compensate for the lower symbol rate, when compared with QOSTBC and CIOD. This results in a worse coding gain.

In addition, we also compare the complexities of ML decoding between these three STBCs. For QOSTBC, $x_1$ and $x_2$ (or $x_3$ and $x_4$) need to be searched jointly. To decode the $2R$ bits in $x_1$ and $x_2$, the number of searches is $2^{2R}$. Hence the total number of searches for decoding the $4R$ bits in $x_1,x_2,x_3,x_4$ is $2^{2R+1}$. For CIOD, it is noted that $s_1$ and $s_2$ are the actual information symbols, and they can be decoded separately. Furthermore, $s_1$ and $s_2$ should be with a higher order modulation to compensate for the loss of symbol rate, when compared with QOSTBC. Therefore, to decode the $2R$ bits in $s_1$, the number of searches is $2^{2R}$. Hence the total number of searches for decoding the $4R$ bits in $s_1$ and $s_2$ is $2^{2R+1}$. We see that these two codes have the same complexity for ML decoding. For OSTBC, as mentioned in the constellation design for (\ref{OSTBC Revised}), every $4R$ bits are divided into three parts, where $2R-1$ bits are mapped to $x_1$, $2R-1$ bits are mapped to $x_2$, and $2$ bits are mapped to $x'_3$. Moreover, $x_1$ and $x_2$ need to be decoded jointly, see (\ref{Precoded OSTBC ML Decoding}). Hence the total number of searches for decoding the $4R$ bits in $x_1,x_2,x'_3$ is ${2^{4R - 2}} + 4$. Obviously, the complexity of OSTBC is much higher than that of QOSTBC and CIOD when $R$ goes high, as shown in Fig. 2.

\subsection{Precoded NZE-TC}

The previously designed precoded OSTBC, precoded QOSTBC, and precoded CIOD just have diversity order of $4$. To achieve a higher diversity order, the codeword matrix with a larger size needs to be considered. However, both of QOSTBC and CIOD are constructed based on OSTBC, therefore the symbol rates of QOSTBC and CIOD are ultimately limited by the symbol rate of OSTBC. The upper bound $3/4$ of the symbol rate of OSTBC with a diversity order higher than $2$ will also be the upper bound of the symbol rates of QOSTBC and CIOD with a diversity order higher than $4$. Hence, both of these two codes suffer from low symbol rates when the diversity order goes high. Moreover, for practical considerations, decoding complexity is an important concern and a decoding scheme with low complexity is always desired. Note that for both of QOSTBC and CIOD, pair-wise ML decoding is required. This may be prohibitive when considering the system implementation, especially for a high-order modulation.  Therefore, STBCs achieving full diversity with a suboptimal receiver that has low decoding complexity, such as a linear receiver, may be favorable.

Many studies have presented STBCs that can achieve full diversity with linear receivers, e.g., \cite{J.Liu2008,Y.Shang2008,W.Zhang2009,H.M.Wang2009}. These codes can achieve full diversity with zero-forcing (ZF) or linear minimum mean-square error (LMMSE) receiver. However, these codeword matrices contain large number of zero entries, therefore they cannot be employed here directly. As mentioned before, zero entries in the low-dimensional codeword matrix make it impossible to have constant power across transmit antennas when with identity matrix precoding. In the meanwhile, the superposition of different symbols after non-identity matrix precoding will complicate the constellation design. Recently, a group of STBCs achieving full diversity with linear receivers and having none zero entry were proposed in \cite{V.B.Pham2015}. The main feature of these codes is that there is no zero entry in the codeword matrices.  Therefore it is possible to have constant power across transmit antennas with these codes. Our design is based on two of these codes, being referred to as no-zero-entry Toeplitz code (NZE-TC) and no-zero-entry overlapped AC (NZE-OAC). The construction of the codeword matrix of NZE-TC is as follows.

Let ${\bf{x}} = {[ {{x_1},{x_2},\ldots,{x_L}} ]}^T$ be the information symbols. The codeword matrix  ${\bf{T}}( {{\bf{x}},L,N} ) \in {\mathbb{C}^{( {L + N - 1} ) \times N}}$ of Toeplitz code is defined as
\begin{align}
{\bf{T}}( {{\bf{x}},L,N} ) = { {\begin{bmatrix}
{{x_1}}&{{x_2}}& \cdots &{{x_L}}&0& \cdots &0\\
0&{{x_1}}&{{x_2}}& \cdots &{{x_L}}& \cdots &0\\
 \vdots & \vdots & \ddots & \ddots & \ddots & \ddots & \vdots \\
0&0& \cdots &{{x_1}}&{{x_2}}& \cdots &{{x_L}}
\end{bmatrix}} ^T} \label{TC}
\end{align}
where
\begin{align}
{[ {{\bf{T}}( {{\bf{x}},L,N} )} ]}_{m,n} = \left\{ {\begin{matrix*}[l]
{{x_{m - n + 1}},}&{n \le m < n + L}\\
{0,}&{\text{other}.}
\end{matrix*}} \right. \nonumber
\end{align}
Then, the codeword matrix ${\bf{X}}_{\mathrm{NT}}( {{\bf{x}},L,N} ) \in {\mathbb{C}^{( {L + N - 1} ) \times N}}$ of NZE-TC is constructed based on (\ref{TC}), which is defined as
\begin{align}
&{\bf{X}}_{\mathrm{NT}}( {{\bf{x}},L,N} ) = { {\begin{bmatrix}
{{x_1}}&{{x_2}}& \cdots &{{x_L}}&{ - {x_1}}& \cdots &{ - {x_{N - 1}}}\\
{{x_L}}&{{x_1}}&{{x_2}}& \cdots &{{x_L}}&{ - {x_1}}& \cdots \\
 \vdots & \ddots & \ddots & \ddots & \ddots & \ddots & \vdots \\
{{x_{L - N + 2}}}& \cdots &{{x_L}}&{{x_1}}&{{x_2}}& \cdots &{{x_L}}
\end{bmatrix}} ^T} \label{NZETC}
\end{align}
where
\begin{align}
&{[ {\bf{X}}_{\mathrm{NT}}( {{\bf{x}},L,N} ) ]}_{m,n} = \left\{ {\begin{matrix*}[l]
{{{[ {{\bf{T}}( {{\bf{x}},L,N} )} ]}_{m + L,n}},}&{m < n}\\
{{{[ {{\bf{T}}( {{\bf{x}},L,N} )} ]}_{m,n}},}&{n \le m < n + L}\\
{ - {{[ {{\bf{T}}( {{\bf{x}},L,N} )} ]}_{m - L,n}},}&{m \ge n + L.}
\end{matrix*}} \right. \nonumber
\end{align}
The symbol rate of NZE-TC is given by
\begin{align}
R = \frac{L}{L+N-1} \label{NZETC Symbol Rate}
\end{align}
since $L$ information symbols are transmitted within $L+N-1$ time slots in (\ref{NZETC}). As long as $L$ is sufficiently large and $N$ is kept as a constant, the symbol rate of NZE-TC will approach $1$.

Here, we consider an arbitrary value for $N$ and $T = L+N-1$ in (\ref{System Model 2}). Correspondingly, the STBC matrix $\bf X$ in (4) is described as the transpose of the NZE-TC matrix in (\ref{NZETC}), i.e., ${({\bf{X}}_{\mathrm{NT}}( {{\bf{x}},L,N} ))}^T$. We have to design the precoding matrix ${\bf{W}}_{\mathrm{NT}} \in {\mathbb{C}^{M \times N}}$ and the constellations of the $L$ information symbols $x_1,x_2,\ldots,x_{L}$ in ${{\bf{X}}_{\mathrm{NT}}( {{\bf{x}},L,N} )}$ to let ${\bf W}_{\mathrm{NT}} {({\bf{X}}_{\mathrm{NT}}( {{\bf{x}},L,N} ))}^T  \in {\mathbb{C}^{M \times (L+N-1)}}$ satisfy Requirements 1--2 simultaneously, i.e., all the $M$ elements in each column of ${\bf F}_M{\bf W}_{\mathrm{NT}} {({\bf{X}}_{\mathrm{NT}}( {{\bf{x}},L,N} ))}^T$ have the same amplitude, and all the $M$ elements in each column of ${\bf W}_{\mathrm{NT}} {({\bf{X}}_{\mathrm{NT}}( {{\bf{x}},L,N} ))}^T$ also have the same amplitude.

With Lemma 3, we let the number of BS antennas $M$ be an integer multiple of $N^2$. Correspondingly, the precoding matrix is proposed to be
\begin{align}
{\bf W}_{\mathrm{NT}} = {\rm{diag}}( {\bf{c}} )( {{{\bf{1}}_{M/N}} \otimes {{\bf{I}}_N}} ) . \label{Precoded NZETC Precoding Matrix}
\end{align}
Therefore, the transmitted signal is ${\bf W}_{\mathrm{NT}} {\bf X}_{\mathrm{NT}} = {\rm{diag}}( {\bf{c}} )\cdot( {{\bf{1}}_{M/N}} \otimes  {({\bf{X}}_{\mathrm{NT}}( {{\bf{x}},L,N} ))}^T )$. Lemma 3 reveals that Requirements 1--2 can be satisfied simultaneously if and only if the $L$ information symbols $x_1,x_2,\ldots,x_L$ in (\ref{NZETC}) have the same amplitude. Therefore, all the constellations of information symbols $x_1,x_2,\ldots,x_L$ are selected to be PSK.
Furthermore, this STBC can achieve diversity order of $N$ with linear receivers \cite{V.B.Pham2015}.

\subsection{Precoded NZE-OAC}

The construction of the codeword matrix of NZE-OAC is based on NZE-TC. First, we rewrite ${\bf{X}}_{\mathrm{NT}}( {{\bf{x}},L,N} )$ in (\ref{NZETC}) as
\begin{align}
{\bf{X}}_{\mathrm{NT}}( {{\bf{x}},L,N} ) = [ {{{\bf{x}}_1},{{\bf{x}}_2},\ldots ,{{\bf{x}}_N}} ] \nonumber
\end{align}
where ${\bf x}_n$ denotes the $n$th column of ${\bf{X}}_{\mathrm{NT}}( {{\bf{x}},L,N} )$. For the case with an odd $N$, we define two matrices
\begin{align}
{{\bf{X}}_{\rm{NT,o}}}( {{\bf{x}},L,N} ) &= [ {{\bf{x}}_1^*,{{\bf{x}}_2}, \ldots ,{\bf{x}}_{N - 2}^*,{{\bf{x}}_{N - 1}},{\bf{x}}_N^*} ] \nonumber \\
{{\bf{X}}_{\rm{NT,e}}}( {{\bf{x}},L,N} ) &= [ {{{\bf{x}}_N}, -{\bf{x}}_{N - 1}^*, \ldots , {{\bf{x}}_3}, -{\bf{x}}_2^*,{{\bf{x}}_1}} ] \nonumber
\end{align}
and two vectors
\begin{align}
{{\bf{x}}_{\rm{o}}} &= {[ {{x_1},0,{x_3},0,\ldots,{x_{L - 1}},0} ]}^T \nonumber \\
{{\bf{x}}_{\rm{e}}} &= {[ {0,{x_2},0,{x_4},\ldots,0,{x_{L}}} ]}^T \nonumber
\end{align}
for an even $L$, where ${{\bf{x}}_{\rm{o}}}$ keeps all the components of $\bf x$ with odd indices and replace the other components by zeros, and correspondingly ${{\bf{x}}_{\rm{e}}}$ instead keeps all the components of $\bf x$ with even indices. Finally, the codeword matrix ${\bf{X}}_{\mathrm{NO}}( {{\bf{x}},L,N} ) \in {\mathbb{C}^{( {L + N - 1} ) \times N}}$ of NZE-OAC is constructed as
\begin{align}
{\bf{X}}_{\mathrm{NO}}( {{\bf{x}},L,N} ) = {{\bf{X}}_{\rm{NT,o}}}( {{{\bf{x}}_{\rm{o}}},L,N} ) + {{\bf{X}}_{\rm{NT,e}}}( {{{\bf{x}}_{\rm{e}}},L,N} )   \label{NZEOAC}
\end{align}
for an odd $N$. Since $L$ information symbols are transmitted within $L+N-1$ time slots, the symbol rate of (\ref{NZEOAC}) is $L/(L+N-1)$. For the case with an even $N$, we consider the codeword matrix ${\bf{X}}_{\mathrm{NO}}( {{\bf{x}},L,N+1} ) \in {\mathbb{C}^{( {L + N} ) \times (N+1)}}$ constructed from (\ref{NZEOAC}) since $N+1$ is odd. If we delete the first column of this matrix, it can be verified that the first and the last row of the yielding $(L+N)\times N$ codeword matrix are linearly correlated with the other rows \cite{V.B.Pham2015}. Hence they can be eliminated from the codeword matrix to increase the symbol rate from $L/(L+N)$ to $L/(L+N-2)$ without destroying the code structure. As a result, the symbol rate of NZE-OAC is given by
\begin{align}
R = \left\{ {\begin{matrix*}[l]
{\frac{L}{L + N - 2},}&{N\text{ is even}}\\
{\frac{L}{L + N - 1},}&{N\text{ is odd.}}
\end{matrix*}} \right.  \label{NZEOAC Symbol Rate}
\end{align}
It can be seen that, either $N$ is even or odd, as long as $L$ is sufficiently large and $N$ is kept as a constant, the symbol rate of NZE-OAC will approach $1$.

Here, we consider an arbitrary value for $N$ and $T = L+N-2$ or $T = L+N-1$ when $N$ is even or odd in (\ref{System Model 2}), respectively. Correspondingly, the STBC matrix $\bf X$ in (4) is described as the transpose of the NZE-TC matrix, i.e., ${({\bf{X}}_{\mathrm{NO}}( {{\bf{x}},L,N} ))}^T$. We have to design the precoding matrix ${\bf{W}}_{\mathrm{NO}} \in {\mathbb{C}^{M \times N}}$ and the constellations of the $L$ information symbols $x_1,x_2,\ldots,x_{L}$ in ${{\bf{X}}_{\mathrm{NO}}( {{\bf{x}},L,N} )}$ to let ${\bf W}_{\mathrm{NO}} {({\bf{X}}_{\mathrm{NO}}( {{\bf{x}},L,N} ))}^T  $ satisfy Requirements 1--2 simultaneously, i.e., all the $M$ elements in each column of ${\bf F}_M{\bf W}_{\mathrm{NO}} {({\bf{X}}_{\mathrm{NO}}( {{\bf{x}},L,N} ))}^T$ as well as all the $M$ elements in each column of ${\bf W}_{\mathrm{NO}} {({\bf{X}}_{\mathrm{NO}}( {{\bf{x}},L,N} ))}^T$ have the same amplitude.

Similar to precoded NZE-TC, with Lemma 3, we let the number of BS antennas $M$ be an integer multiple of $N^2$. Correspondingly, the precoding matrix is proposed to be
\begin{align}
{\bf W}_{\mathrm{NO}} = {\rm{diag}}( {\bf{c}} )( {{{\bf{1}}_{M/N}} \otimes {{\bf{I}}_N}} ) . \label{Precoded NZEOAC Precoding Matrix}
\end{align}
Therefore, the transmitted signal is ${\bf W}_{\mathrm{NO}} {\bf X}_{\mathrm{NO}} = {\rm{diag}}( {\bf{c}} )\cdot( {{\bf{1}}_{M/N}} \otimes  {({\bf{X}}_{\mathrm{NO}}( {{\bf{x}},L,N} ))}^T )$. Lemma 3 reveals that Requirements 1--2 can be satisfied simultaneously if and only if the $L$ information symbols $x_1,x_2,\ldots,x_L$ in (\ref{NZEOAC}) have the same amplitude. Therefore, all the constellations of information symbols $x_1,x_2,\ldots,x_L$ are selected to be PSK. Furthermore, this STBC can achieve diversity order of $N$ with linear receivers \cite{V.B.Pham2015}.

\section{Numerical Results}

In this section, we present numerical simulations to evaluate the performance of the proposed omnidirectional STBCs for common information broadcasting in massive MIMO systems. We consider a 120$^\circ$ sector. The BS is with a ULA of $M = 128$ antennas, where the antenna space is $d = \lambda / \sqrt{3}$. In Section IV-B we have shown that the number of UTs $K$ does not affect the diversity order. Therefore we let $K = 1$ here. The UT has one single antenna and may have different positions within the sector. The channel between the BS and the UT is modeled according to (\ref{Channel Model 0}) and (\ref{Channel Model 1}). We consider the typical outdoor propagation environments where the PAS in (\ref{Channel Model 1}) is assumed to follow truncated Gaussian distribution \cite{Y.S.Cho2010}
\begin{align}
p( \theta  ) = {\exp \bigg( { - \frac{{{{( {\theta  - {\theta _0}} )^2}}}}{{2{\sigma ^2}}}} \bigg) , } \;\; { { - \frac{\pi}{2} \leq \theta \leq \frac{\pi}{2}} } \label{PAS}
\end{align}
where $\theta _0$ and $\sigma $ denote the mean angle of departure (AoD) and the angle spread (AS), respectively, and the subscript index $k$ has been omitted since there is only $K=1$ UT. We let $\sigma = 5^{\circ}$ and let $\theta _0 $ vary in $[ { - \pi /3,\pi /3} ]$ to represent the cases when the UT has different angles with respect to the ULA of the BS within the sector. Different values for the noise variance  $\sigma^2_{\mathrm{n}}$ in (\ref{System Model 2}) are used to represent different distances between the BS and the UT, while the total average transmission power at the BS and the large-scale fading coefficient of the channel are normalized to unit.

We evaluate the bit error rate (BER) performance of the STBC designs proposed in Section V, including: 1) single-stream precoding, 2) precoded AC, 3) precoded OSTBC, 4) precoded QOSTBC, 5) precoded CIOD, 6) precoded NZE-TC, and 7) precoded NZE-OAC. Bit rates of $1$ and $2$ bps/Hz are considered. For precoded OSTBC, as mentioned in the constellation design for (\ref{OSTBC Revised}), we let $x_1 \in \mathcal{S}_{\mathrm{BPSK}}$, $x_2 \in j \mathcal{S}_{\mathrm{BPSK}}$, and $x'_3 \in \mathcal{S}_{\mathrm{QPSK}}$ for $1$ bps/Hz, and let $x_1 \in \mathcal{S}_{8\mathrm{PAM}}$, $x_2 \in j \mathcal{S}_{8\mathrm{PAM}}$, and $x'_3 \in \mathcal{S}_{\mathrm{QPSK}}$ for $2$ bps/Hz, respectively. For precoded NZE-TC and precoded NZE-OAC, it is difficult to let their bit rates be equal to $1$ or $2$ bps/Hz strictly, see (\ref{NZETC Symbol Rate}) and (\ref{NZEOAC Symbol Rate}). Therefore, we let $L = 30$ and $N = 8$ in (\ref{NZETC Symbol Rate}) and (\ref{NZEOAC Symbol Rate}), yielding an $8 \times 37$ codeword matrix with symbol rate $30/37 \approx 0.81$ for NZE-TC, and an $8 \times 36$ codeword matrix with symbol rate $30/36 \approx 0.83$ for NZE-OAC, respectively. The corresponding bit rates are $0.81$ bps/Hz or $1.62$ bps/Hz for NZE-TC, and $0.83$ bps/Hz or $1.66$ bps/Hz for NZE-OAC, respectively, when the information symbols therein are with BPSK or QPSK constellation. The other STBCs use BPSK or QPSK constellation directly, yielding bit rates of $1$ or $2$ bps/Hz. All the roots of the ZC sequences $\bf c$ in precoding matrices (\ref{Precoding Vector}), (\ref{Precoded Alamouti Code Precoding Matrix}), (\ref{Precoded OSTBC Precoding Matrix}), (\ref{Precoded QOSTBC Precoding Matrix}), (\ref{Precoded CIOD Precoding Matrix}), (\ref{Precoded NZETC Precoding Matrix}), and (\ref{Precoded NZEOAC Precoding Matrix}) are set as $\gamma = 1$. The effective channel ${\bf W}^H{\bf h}$ is assumed to be perfectly known at the UT side. ZF receiver is used for precoded NZE-TC and precoded NZE-OAC, while ML receiver is used for the others.

\begin{figure}[htbp]
\centering
\includegraphics[width=0.75\columnwidth]{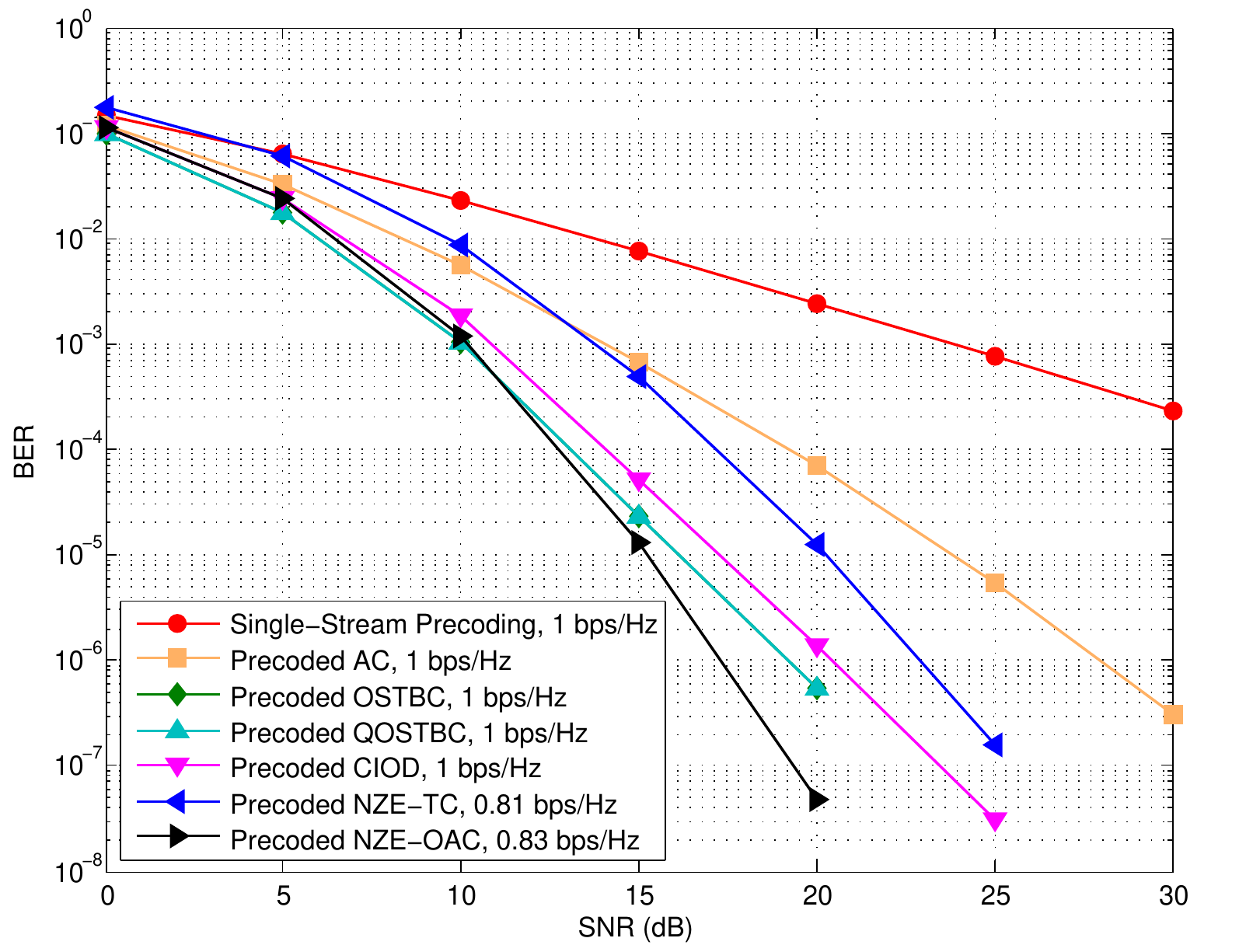}
\caption{BER performance comparison between different designs for $1$ bps/Hz. Results are shown versus the SNR value.}
\end{figure}

\begin{figure}[htbp]
\centering
\includegraphics[width=0.75\columnwidth]{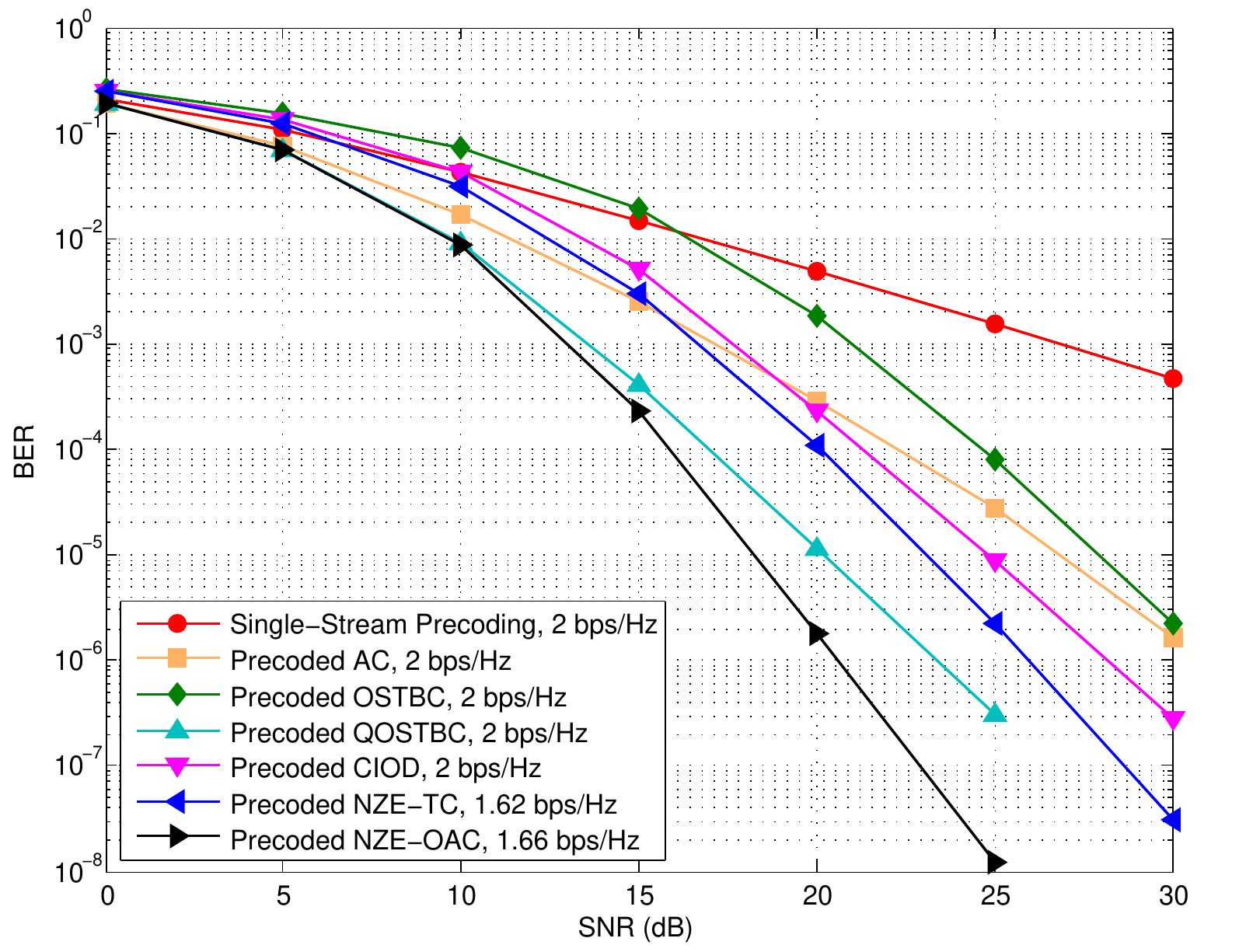}
\caption{BER performance comparison between different designs for $2$ bps/Hz. Results are shown versus the SNR value.}
\end{figure}

\begin{figure}[htbp]
\centering
\includegraphics[width=0.75\columnwidth]{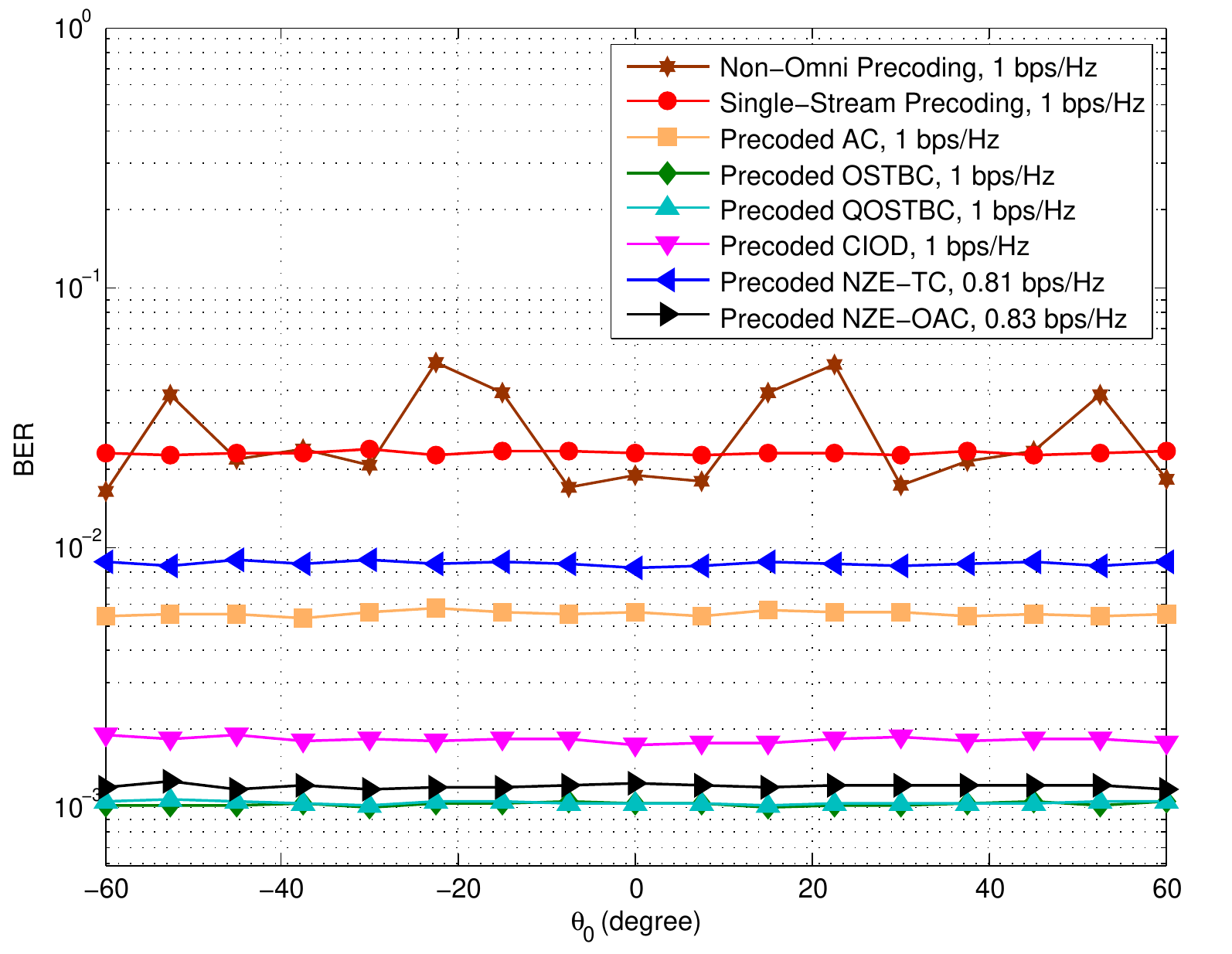}
\caption{BER performance comparison between different designs for $1$ bps/Hz and SNR $= 10$ dB. Results are shown versus the mean AoD $\theta_0$.}
\end{figure}

First, we evaluate the BER performance to verify the diversity performance, where the mean AoD in (\ref{PAS}) is fixed as $\theta_0 = 0$. The BER performance with respect to the SNR value, i.e., $1/\sigma^2_{\mathrm{n}}$, for $1$ and $2$ bps/Hz for the above STBCs is presented in Fig. 3 and Fig. 4, respectively. The slopes of the BER curves at high SNR values indicate the corresponding diversity orders of these designs, where single-stream precoding has diversity order of $1$, precoded AC has diversity order of $2$, precoded QOSTBC and precoded CIOD have diversity order of $4$, and the other two STBCs have diversity order of $8$. This verifies the analytical results in Section IV-B. Moreover, the BER performance of precoded QOSTBC is the same with that of precoded OSTBC for $1$ bps/Hz. This is because they have the same coding gain for $1$ bps/Hz as in Fig. 1. Similarly, precoded CIOD underperforms precoded QOSTBC and precoded OSTBC for $1$ bps/Hz. This also verifies the results in Fig. 1. Additionally, the above results of the much worse performance of precoded NZE-TC than that of precoded NZE-OAC here agree with the results in the conventional small-scale MIMO systems \cite{Y.Shang2008}.

Then, we evaluate the BER performance to verify the ability of omnidirectional transmission for the STBC designs, which means that when the distance between the BS and the UT is fixed, the BER performance keeps constant with respect to the angle between the UT and the ULA of the BS. The BER performance with respect to the mean AoD $\theta_0$ in (\ref{PAS}) for $1$ bps/Hz and SNR $ = 10 $ dB is presented in Fig. 5. For each value of $\theta_0$, the spatial covariance matrix $\bf R$ is generated from (\ref{Channel Model 1}) and (\ref{PAS}), and the channel vector $\bf h$ follows Rayleigh fading as in (\ref{Channel Model 0}). We observe that all the proposed designs show constant BER performance for different values of $\theta_0$ when the SNR is fixed. As a comparison, if the omnidirectional transmission criterion in Requirement 1 is not satisfied, e.g., using a pseudo-random binary sequence instead of the ZC sequence as the precoding vector $\bf w$ in (\ref{Precoding Vector}), the corresponding BER performance may not be constant with respect to $\theta_0$. This is shown as the ``non-omni precoding'' curve in Fig. 5.

As a final remark, since any STBC of a proper dimension can be directly applied in \cite{X.Meng2016} without the need of any special designs of the signals, where only constant average power is required, the BER performance and the decoding complexity of the precoded STBC in this paper may be worse and higher than that in \cite{X.Meng2016}, respectively. Take precoded AC for example. If constant instantaneous power is required, in Section V-B we have shown that only PSK can be applied for the information symbols in (\ref{Alamouti Code}). Nevertheless, if only constant average power is required, more efficient QAM can be applied for the information symbols in (\ref{Alamouti Code}). It is known that QAM is superior to PSK in terms of both BER performance and decoding complexity. This is because QAM has a larger minimum Euclidean distance between different constellation points than PSK for a fixed size, and the real and imaginary parts in QAM can be decoded separately, while in PSK they must be decoded jointly.

\section{Conclusions}

In this paper, the approach in terms of space-time block coding was investigated to broadcast the common information omnidirectionally in a massive MIMO downlink. Since the direct estimation of the actual channel between a massive antenna array of a BS and a UT takes too many time-frequency resources and hence significantly lowers the net spectral efficiency, the high-dimensional STBC transmitted from the BS antennas was proposed to be composed by a channel-independent precoding matrix and a low-dimensional STBC. Consequently, just the effective low-dimensional channel after precoding needs to be estimated at the UT side and the corresponding pilot overhead can be significantly reduced, and at the same time, partial transmit diversity can be achieved. Under this framework, a systematic approach was proposed to jointly design the precoding matrix and the signal constellation in the low-dimensional STBC, to guarantee omnidirectional transmission, equal instantaneous power on each transmit antenna, and at the same time, achieve the full diversity of the low-dimensional STBC. Finally, several detailed examples were designed based on classic low-dimensional STBCs.

\appendices

\section{Proof of Lemma 3}

Let ${\bf{x}} = {[ {{x_0}, {x_1}, \ldots ,{x_{N - 1}}} ]}^T$ and ${\bf c} = \frac{1}{\sqrt{M}}\cdot{[c_0,c_1,\ldots,c_{M-1}]}^T $. Then we have
\begin{align}
&{\rm{diag}}( {\bf{c}} )( {{{\bf{1}}_{M/N}} \otimes {\bf{x}}} ) = \frac{1}{\sqrt{M}}  \cdot[{c_0x_0}, \ldots ,{c_{N-1}x_{N - 1}},{\ldots ,{c_{M-N}x_0}, \ldots ,{c_{M-1}x_{N - 1}}]}^T . \nonumber
\end{align}
Since $c_0,c_1,\ldots,c_{M-1}$ in the ZC sequence $\bf c$ have the same amplitude, it is straightforward to see that all the $M$ elements in ${\rm{diag}}( {\bf{c}} )( {{{\bf{1}}_{M/N}} \otimes {\bf{x}}} )$ have the same amplitude if and only if $x_0,x_1,\ldots,x_{N-1}$ have the same amplitude.

Then, we need to prove that all the $M$ elements in the $M$-point DFT of ${\rm{diag}}( {\bf{c}} )( {{{\bf{1}}_{M/N}} \otimes {\bf{x}}} )$ have the same amplitude if and only if $x_0,x_1,\ldots,x_{N-1}$ have the same amplitude. It can be verified that the normalized $M$-point DFT of ${{\bf{1}}_{M/N}} \otimes {\bf{x}}$ can be expressed as
\begin{align}
& {{\bf{1}}_{M/N}} \otimes {\bf{x}}  = {[{x_0}, \ldots ,{x_{N - 1}}, \ldots ,{x_0}, \ldots ,{x_{N - 1}}]}^T  \nonumber \\
& \xrightarrow{\text{DFT}} \sqrt {\frac{M}{N}} \cdot {[{X_0},\underbrace {0, \ldots ,0}_{\frac{M}{N} - 1\text{ times}}, \ldots ,{X_{N - 1}},\underbrace {0, \ldots ,0}_{\frac{M}{N} - 1\text{ times}}]}^T \label{Lemma 1 Proof 1}
\end{align}
where
\begin{align}
{X_k} = \frac{1}{{\sqrt N }}\sum\limits_{n = 0}^{N - 1} {{x_n}{e^{ - j2\pi kn/N}}} , \;\; k = 0,1, \ldots ,N - 1. \label{Lemma 1 Proof 2}
\end{align}
With the results in \cite{B.M.Popovic2010}, the normalized $M$-point DFT of $\bf{c}$ defined in (\ref{ZC Sequence}) can be expressed as
\begin{align}
&{C_k} = \frac{1}{{ M }}\sum\limits_{m = 0}^{M - 1} {{c_n}{e^{ - j2\pi km/M}}} = c  {e^{ - j\pi \gamma {\gamma ^{ - 1}}k({\gamma ^{ - 1}}k - {{((M))}_2})/M}},\;\;k = 0,1, \ldots ,N - 1 \label{Lemma 1 Proof 3}
\end{align}
where $c = \frac{1}{M} \sum\nolimits_{m = 0}^{M - 1} {{c_m}} $ and ${\gamma}^{-1}$ is a positive integer less than $M$ and satisfies ${(( \gamma {\gamma ^{ - 1}} ))}_M = 1$.

Let $B_k$ be the $k$th element of the normalized $M$-point DFT of ${\rm{diag}}( {\bf{c}} )( {{{\bf{1}}_{M/N}} \otimes {\bf{x}}} )$. Since ${\rm{diag}}( {\bf{c}} )( {{{\bf{1}}_{M/N}} \otimes {\bf{x}}} )$ can be seen as the element-wise multiplication of ${\bf{c}}$ and ${{{\bf{1}}_{M/N}} \otimes {\bf{x}}}$, with the property of DFT, i.e., element-wise multiplication in the time domain is equivalent to cyclic convolution in the frequency domain, we have
\begin{align}
{B_k} &= \frac{1}{{\sqrt N }}\sum\limits_{n = 0}^{N - 1} {{C_{{((k - Mn/N))}_M}}{X_n}} \nonumber \\
&= \frac{1}{N}\sum\limits_{l,n = 0}^{N - 1} {c{e^{ - j\pi \gamma {\gamma ^{ - 1}}(k - Mn/N)({\gamma ^{ - 1}}(k - Mn/N) - {{((M))}_2})/M}}} {x_l}{e^{ - j2\pi ln/N}} \nonumber \\
&= \frac{{C_k}}{N}\sum\limits_{l,n = 0}^{N - 1} {{x_l}{e^{j2\pi f(k,l)n/N}}}, \;\; k = 0, 1, \ldots ,N - 1 \label{Lemma 1 Proof 4}
\end{align}
where the first equality follows from (\ref{Lemma 1 Proof 1}), the second equality is due to (\ref{Lemma 1 Proof 2}) and (\ref{Lemma 1 Proof 3}), and the last equality is based on the fact that ${e^{ - j\pi \gamma {{({\gamma ^{ - 1}})}^2}M{n^2}/{N^2}}} = {e^{ - j\pi \gamma {{({\gamma ^{ - 1}})}^2}Mn/{N^2}}}$ for any integer $n$ when $M$ is an integer multiple of $N^2$ and
\begin{align}
f( {k,l} ) \triangleq {\Bigg( \!\!{\Bigg( {\gamma {{({\gamma ^{ - 1}})}^2}k - l - \frac{{M\gamma {{({\gamma ^{ - 1}})}^2}}}{{2N}} - \frac{{{{((M))}_2}\gamma {\gamma ^{ - 1}}}}{2}} \Bigg)}\!\! \Bigg)\!\!}_N. \nonumber
\end{align}
It is not hard to verify that, for any $k \in \{ {0,1, \ldots ,M - 1} \}$, there is one and only one $l \in \{ {0,1, \ldots ,N - 1} \}$ could let $f( {k,l} ) = 0$. We denote this $l$ as $l( k )$, i.e., $f( {k,l( k )} ) = 0$. Therefore, we can express (\ref{Lemma 1 Proof 4}) as
\begin{align}
{B_k} = {C_k}{x_{l(k)}}. \label{Lemma 1 Proof 5}
\end{align}

The function ${l( k )}$ is defined in the domain that $k \in \{ {0,1, \ldots ,M - 1} \}$ and takes values in $\{ {0, 1, \ldots ,N - 1} \}$. It has the following two properties:
\begin{itemize}
\item ${l( k )}$ is a periodic function with minimum period $N$, i.e., $l( k ) = l( {k + N} )$;
\item $l( k ) \ne l( {k'} )$ for ${( {( k )} )}_N \ne {( {( {k'} )} )}_N$.
\end{itemize}
With these two properties and (\ref{Lemma 1 Proof 5}), we have
\begin{align}
{[ {{B_0},{B_1}, \ldots , {B_{M - 1}}} ]}^T &= {\rm{diag}}\{ {{C_0},{C_0}, \ldots ,{C_{M - 1}}} \}({{\bf{1}}_{M/N}} \otimes {[ {{x_{l( 0 )}}, {x_{l( 1 )}}, \ldots ,{x_{l( {N - 1} )}}} ]}^T) . \nonumber
\end{align}
Hence, if and only if ${x_{l( 0 )}}, {x_{l( 1 )}}, \ldots ,{x_{l( {N - 1} )}}$ have the same amplitude, which is equivalent to ${x_{0}}, {x_{1}}, \ldots ,{x_{{N - 1} }}$ having the same amplitude, ${B_0}, B_1,\ldots ,{B_{M - 1}}$ will have the same amplitude since $C_0,C_1,\ldots,C_{M-1}$ have the same amplitude. Therefore, all the $M$ elements in the $M$-point DFT of ${\rm{diag}}( {\bf{c}} )( {{{\bf{1}}_{M/N}} \otimes {\bf{x}}} )$ have the same amplitude if and only if ${x_{0}}, {x_{1}}, \ldots ,{x_{{N - 1} }}$ have the same amplitude.

Finally, we conclude that ${\rm{diag}}( {\bf{c}} )( {{{\bf{1}}_{M/N}} \otimes {\bf{x}}} )$ is a CAZAC sequence of length $M$, if and only if ${x_{0}}, {x_{1}}, \ldots ,{x_{{N - 1} }}$ have the same amplitude.

\end{document}